\documentclass[review]{elsarticle}

\usepackage{lineno,hyperref}

\usepackage{amsmath}
\usepackage{amssymb}
\usepackage{booktabs}
\usepackage{caption}
\usepackage{graphicx}
\usepackage{float}
\usepackage{subcaption}

\begin{document}

\begin{frontmatter}
\hypersetup{linkcolor = blue,anchorcolor =red,citecolor = blue,filecolor = red,urlcolor = red,
            pdfauthor=author}


\title{Echo chamber effects based on a novel three-dimensional Deffuant-Weisbuch model\tnoteref{t1,t2}}
\tnotetext[t1]{\text{AMS Subject Classification}: 91D30 \sep 05C82 \sep 05C80 \sep 05C85.}
\tnotetext[t2]{This work is supported by the State Key Program of National Natural Science of China under Grant No.91324201. This work is also supported by the Fundamental Research Funds for the Central Universities of China under Grant 2018IB017, Equipment Pre-Research Ministry of Education Joint Fund Grant 6141A02033703 and the Natural Science Foundation of Hubei Province of China under Grant 2014CFB865.
}


\author[mymainaddress]{Fei Gao\corref{mycorrespondingauthor}
\author[mainaddress]{Yuxin Xu}}
\ead{gaof@whut.edu.cn}


\cortext[mycorrespondingauthor]{Corresponding author.}


\address[mymainaddress]{Department of Statistics and Center for Mathematical Sciences, Wuhan University of Technology, Wuhan, 430070, China}
\address[mainaddress]{Department of Mathematics and Center for Mathematical Sciences, Wuhan University of Technology, Wuhan, 430070, China}

\begin{abstract}
In order to solve the problem of opinion polarization and distortion caused by echo chamber effect in the evolution process of online public opinion, a novel three-dimensional Deffuant-Weisbuch model is proposed to study the formation and elimination of echo chamber effect in this paper. Firstly, the original pairwise interaction model is generalized to three-point interaction model. Secondly, we consider individual psychological mechanism and introduce individual emotional factor into the trust threshold of original model. Finally, the natural evolution coefficient of opinion is introduced to modify the model. The improved model is used to conduct simulation experiments on social networks with different structures, and opinion leaders and active agents are introduced into the network, so as to study the corresponding generation and breaking mechanism of echo chamber. The experimental results show that the change of network structure cannot eliminate the echo chamber effect, and the increase of network stability and connectivity can only slow down the echo chamber effect. Opinion leaders can aggregate opinions within their scope of influence and have a guiding effect on opinions. Therefore, if opinion leaders can change their opinions over time, they can well guide opinions to converge to neutral opinions, thus achieving the purpose of breaking the echo chamber. Active agents can lead the opinions in the network to converge to the neutral, and active agents with high stubbornness can lead the free views to converge to the neutral, thus achieving the purpose of breaking the echo chamber effect.
\end{abstract}

\begin{keyword}
 Social network \sep Opinion dynamics \sep Echo chamber effect \sep three-dimensional Deffuant-Weisbuch model
\MSC[2010] 00-01\sep  99-00
\end{keyword}

\end{frontmatter}


\section{Introductions}
Nowadays, with the highly developed Internet and the unprecedented speed of information transmission, online public opinion events often occur , echo chamber effect widely exists in the process of information transmission, reception and interaction. In China, the concept of echo chamber effect was first defined by Hu Yong\cite{1}. which was defined as a modern communication phenomenon caused by the continuous dissemination, exaggeration and distortion of similar information in a relatively closed opinion perception environment. After the concept of echo chamber effect was put forward, some scholars verified the existence of echo chamber effect from different point of view, Bruns et al.\cite{2} studied the evidence of the existence of echo chamber effects in their country through the interaction between 4 million Twitter accounts. Wang et al. \cite{3} used network analysis to prove that during the epidemic period, users did show obvious echo chamber effect when forwarding or commenting rumor refutation information with different authenticity. Cota et al. \cite{4} demonstrated the existence of echo chamber effect in political communication networks. and some scholars also established different models to analyze the causes of echo chamber effect. Bessi\cite{5} analyzed the influence of users` psychological factors and individual personality characteristics on the formation of echo chamber effect, and believed that the existence of specific personality characteristics of users directly led to their participation in the formation of echo chamber. Matz et al. \cite{6} also discussed the influence of agents` personality factors on the formation of echo chamber in their paper. Vicario et al. \cite{7} used three growth models to discuss the influence of users' emotional factors on the formation of echo chamber effect.

In many real situations, the existence of echo chamber effect often leads to adverse phenomena such as amplification of extremism and dissemination of false information \cite{8}. Also it is very difficult to reach a complete consensus when online public opinion events occur because there are always two or several inverse opinions. As time goes by, agents with similar opinions will gather together and will be isolated from heterogeneous opinions, eventually forming an echo chamber \cite{9}. More and more empirical evidence prove this phenomenon: On the one hand, polarization of opinions is found in social media dialogues, which is extremely obvious in the political field of partisanship \cite{10}. The disagreement of political opinions will lead to less and less communication between parties, resulting in polarization of views within parties \cite{11}. On the other hand, low diversity is found in online information consumption, and people will only spread relevant information among homogeneous individuals\cite{12}.The phenomena above all have one thing in common, that is, heterogeneous group isolation ,it can be well explained by "Birds of a feather flock together".

The echo chamber effect widely exists in the process of public opinion evolution and opinion interaction, therefore, most opinion dynamics models \cite{13} are suitable for studying the echo chamber effect, including bounded confidence model \cite{14,15,16}, voter model \cite{17,18}, Sznajd model \cite{19,20}, DeGroot model \cite{21,22}, binary naming game method \cite{23,24} and so on. Among them, the bounded confidence model is widely used because it is more in line with the actual interaction process of opinions. Bounded confidence models include Deffuant-Weisbuch model \cite{25} and Hegselmann-Krause model \cite{15}, both of which are continuous opinions interaction models. Before that, the pairwise D-W model has been widely used to study the evolution of online public opinions. Zhang et al. \cite{26} studied the acceptance of agent`s opinions and the influence of opinion leaders on the evolution of online public opinion. Sobkowicz \cite{27} considered the psychological elements and emotional factors of each agent in reality, modified the trust threshold in the model, and introduced emotional factors to make the model more in line with the actual interaction process of opinions. Malarz et al. \cite{28}introduced Zaller model of public opinion into the D-W model. Luo et al. \cite{29} introduced implicit and explicit opinions on the basis of the original D-W model, and believed that each agent should have two opinions. The above-mentioned D-W models and their improvements all have a common feature, namely pairwise interaction. However, in many real situations, the interaction of opinions does not only occur between two people. Specifically, in social networks, the interaction of opinions may occur between a node and its neighbor nodes. Therefore, the pairwise interaction model often fails to reflect the real situation of the interaction of opinions.

Here, we propose a three-dimensional D-W model based on the 2D D-W model, that is, from pairwise interaction to three-point interaction, so as to be closer to the reality. In the original pairwise D-W model, it is usually assumed that only one opinion interaction will occur within a time step. However, in this paper, there is no limit on the number of opinion interaction within a time step to ensure that nodes can complete the interaction with qualified nodes within a time step. At the same time, the natural evolution coefficient of opinions was introduced into the model \cite{30} to achieve the purpose of model modification. In addition, we improve the trust threshold according to agent’s psychological factors and emotional factors, and combined with the natural evolution coefficient of opinion. Different agents have different trust thresholds.

The rest of the paper is organized as follows. In the second section, the proposed model and the corresponding parameter improvement will be introduced in detail. In the third section, we will use social networks with different structures and introduce opinion leaders and active individuals into the network to conduct simulation experiments. Conclusions will be drawn in the last section.

\section{A novel three dimensional Deffuant-Weisbuch model}
\subsection{Deffuant-Weisbuch model}
The D-W model\cite{25} is an opinion dynamics model based on agents’ interactions established by Weisbuch and Deffuant. The model assumes that the interaction of agents’ opinions is affected by their trust threshold, and the interaction of opinions will occur only when both opinions are within the trust threshold. The D-W model sets agent opinions to any value within , different from Sznajd model\cite{17}, voter model\cite{19} and other discrete models, D-W model depicts the interaction of agents’ opinions from a continuous perspective. In addition, in the D-W model, the interaction of opinions between agents occurs randomly. At each time step, two random agents of the group interact according to the following rules: Interaction can occur if the differences between their opinions are less than a given threshold , otherwise the interaction will not occur.

According to the original form of the D-W model, it is assumed that the number of agents is N, and the trust threshold is $ \varepsilon  \in [0,1] $, In each time step $t$, agents $i$ and $j$ are randomly selected from the set of agents to interact with each other. Let the opinion values of $i$ and $j$ be ${o_i}(t)$ and ${o_j}(t)$ respectively, and ${o_i}(t),{o_j}(t) \in [0,1]$ ,if $\left| {{o_i}(t) - {o_j}(t)} \right| \le \varepsilon$  ,then:
\begin{equation}
    \left\{ \begin{array}{l}
{o_i}(t + 1) = {o_i}(t) + \mu ({o_j}(t) - {o_i}(t))\\
{o_j}(t + 1) = {o_j}(t) + \mu ({o_i}(t) - {o_j}(t))
\end{array} \right.
\end{equation}

Otherwise:
\begin{equation}
  \left\{ \begin{array}{l}
{o_i}(t + 1) = {o_i}(t)\\
{o_j}(t + 1) = {o_j}(t)
\end{array} \right.
\end{equation}

Parameter$\mu$is very important in the D-W model, and groups with$\mu$ different properties can be obtained by adjusting parameter values $\mu$. Generally speaking, the value of parameter is between $[0,0.5]$. When $\mu  = 0$, its value represents extremely stubborn agents whose opinions will not change due to interaction. When $\mu  = 0.5$, its value represents the compromised agents, and the two sides of the interaction of opinions update their respective opinions with their mean values. To a certain extent, the value also reflects the degree of stubbornness of agents to their own opinions. When the value of $\mu$ is greater than 0.5, it indicates that the opinions of the two sides in the interaction tend to exchange and they are more inclined to choose the opinions of the other side, which rarely happens in real life, so the value of is usually limited between $[0,0.5]$. In fact, in the original D-W model, $\mu$ value is usually set as 0.5 to simplify the model, so the corresponding update rule of opinion value becomes:
\begin{equation}
\left| {{o_i}(t) - {o_j}(t)} \right| \le \varepsilon  \to \left\{ \begin{array}{l}
{o_i}(t + 1) = {o_i}(t) + 0.5({o_j}(t) - {o_i}(t))\\
{o_j}(t + 1) = {o_j}(t) + 0.5({o_i}(t) - {o_j}(t))
\end{array} \right.
\end{equation}

\begin{equation}
\left| {{o_i}(t) - {o_j}(t)} \right| > \varepsilon  \to \left\{ \begin{array}{l}
{o_i}(t + 1) = {o_i}(t)\\
{o_j}(t + 1) = {o_j}(t)
\end{array} \right.
\end{equation}

Where the value of trust threshold $\varepsilon $ will greatly affect the process of opinion evolution. When  $\varepsilon  \ge 0.5$, the group tends to form a consistent opinion. With the continuous decrease of  $\varepsilon $, the group will gradually divide into two or even more opinion groups, and each opinion group holds the same opinion, that is, the size of the trust threshold is roughly inversely proportional to the number of formed opinion group.
\subsection{The three dimensional Deffuant-Weisbuch model}
In the original D-W model, we consider the interaction between two random agents. In this paper, we will consider the interaction between three agents. Therefore, based on the original D-W model, the three dimensional D-W model is proposed. Similarly, assuming that the number of agents is N, the value of trust threshold is still $\varepsilon  \in [0,1]$, agents $i$ , $j$ and $k$ are randomly selected for interaction within each time step $t$, and the opinion values of agents $i$ , $j$ and $k$ are respectively set as ${o_i}(t)$ , ${o_j}(t)$ and ${o_k}(t)$,and ${o_i}(t),{o_j}(t),{o_j}(t) \in [ - 1,1]$. If  $\left| {{o_i}(t) - {o_j}(t)} \right| \le \varepsilon $,  $\left| {{o_k}(t) - {o_j}(t)} \right| \le \varepsilon $ and $\left| {{o_i}(t) - {o_k}(t)} \right| \le \varepsilon $ satisfy, then:

\begin{equation}
\left\{ \begin{array}{l}
{o_i}(t + 1) = {o_i}(t) + {\mu _1}({o_j}(t) - {o_i}(t)) + {\mu _2}({o_k}(t) - {o_i}(t))\\
{o_j}(t + 1) = {o_j}(t) + {\lambda _1}({o_i}(t) - {o_j}(t)) + {\lambda _2}({o_k}(t) - {o_j}(t))\\
{o_k}(t + 1) = {o_k}(t) + {\eta _1}({o_i}(t) - {o_k}(t)) + {\eta _2}({o_j}(t) - {o_k}(t))
\end{array} \right.
\end{equation}

Otherwise:
\begin{equation}
\left\{ \begin{array}{l}
{o_i}(t + 1) = {o_i}(t)\\
{o_j}(t + 1) = {o_j}(t)\\
{o_k}(t + 1) = {o_k}(t)
\end{array} \right.
\end{equation}

Where ${\mu _1} + {\mu _2}$ is between  $[0,0.5]$, and so is ${\lambda _1} + {\lambda _2}$ and  ${\eta _1} + {\eta _2}$.
\subsection{The interaction pattern of opinions}
At each time step $t$, agents are randomly selected from the group, and two neighbors are selected from their neighbor  $ N = \{ {i_1},{i_2},{i_3}, \cdots ,{i_m}\} $ to interact with each other. The update rules of agent $i$'s opinion value are as follows:
\begin{equation}
\begin{array}{c}
o_i^{{t_1}}(t) = f({o_i}(t),{o_{{i_1}}}(t),{o_{{i_2}}}(t)),\\
o_i^{{t_2}}(t) = f(o_i^{{i_1}}(t),{o_{{i_1}}}(t),{o_{{i_3}}}(t)),\\
 \vdots \\
o_i^{{t_n}}(t) = f(o_i^{{t_{n - 1}}}(t),{o_{{i_a}}}(t),{o_{{i_b}}}(t)),
\end{array}
\end{equation}

Where $m$ represents the number of neighbor nodes, $n$represents the number of opinions updated within a time step,  $N = \{ {i_1},{i_2},{i_3}, \cdots ,{i_m}\} $represents the random sequence of neighbor nodes,  $t = \{ {t_1},{t_2},{t_3}, \cdots ,{t_n}\} $represents the set of each small step in a time step,  ${i_a}$ and  ${i_b}$ represent two different neighbors of agent $i$, namely $a \ne b$.

Every time the opinion value of agent is updated, the opinion value of the two neighbors selected in this interaction should also be updated. Only when the opinion value of three agents is updated, an opinion interaction is completed.

Next, the psychological mechanism of the agent in the interaction process and the evolution mechanism of the agent's own opinions over time are further considered to improve the opinion interaction pattern of the three dimensional D-W model:

(1)	Trust threshold: Set agent $i$ has a trust threshold ${\varepsilon _i}(t) \in [0.2,0.4]$. When it interacts with its neighbors  ${o_{{i_1}}}(t)$ and  ${o_{{i_2}}}(t)$, it meets the following opinion update rule $f$:
\begin{equation}
\begin{array}{c}
\left| {{o_i}(t) - {o_j}(t)} \right| \le {\varepsilon _i}(t) \cap \left| {{o_k}(t) - {o_j}(t)} \right| \le {\varepsilon _i}(t) \cap \left| {{o_k}(t) - {o_i}(t)} \right| \le {\varepsilon _i}(t) \to \\
o_i^{12}(t) = f({o_i}(t),{o_{{i_1}}}(t),{o_{{i_2}}}(t))
\end{array}
\end{equation}

Otherwise:
\begin{equation}
o_i^{12}(t) = o_i^{}(t)
\end{equation}

Similarly, the opinion values of two neighbors can be updated.

Inspired by \cite{27}, there is an obvious positive correlation between the emotional degree of agents and the content of opinions, and the interaction range of opinions corresponding to extreme agents is smaller than that of rational agents. In this model,  ${\varepsilon _i}(t)$ is set as a function of the absolute value of opinion$\left| {{o_i}(t)} \right|$, that is:
\begin{equation}
{\varepsilon _i}(t) = g(\left| {{o_i}(t)} \right|) = {\varepsilon _{\min }} + \left| {{o_i}(t)} \right|({\varepsilon _{\max }} - {\varepsilon _{\min }})
\end{equation}
Where  ${\varepsilon _{\max }}$ is the agent's maximum trust threshold, which  is set as  ${\varepsilon _{\max }} = 0.4$ in the study.   ${\varepsilon _{\min}}$ is the minimum trust threshold of an agent, which is set as  ${\varepsilon _{\min }} = 0.2$ in the study.

Based on the above description, update rules for three agents in an interaction can be described as follows:suppose agent $i$ is selected, and its two neighbors are arbitrarily chosen as $j$ and $k$, When the trust threshold conditions  $\left| {{o_i}(t) - {o_j}(t)} \right| \le {\varepsilon _i}(t)$ , $\left| {{o_k}(t) - {o_j}(t)} \right| \le {\varepsilon _i}(t)$ , $\left| {{o_k}(t) - {o_i}(t)} \right| \le {\varepsilon _i}(t)$ are met at the same time, the update rules of the opinion values are as follows:
\begin{equation}
    \left\{ \begin{array}{c}
{o_i}^{jk}(t) = (1 - {\mu _1} - {\mu _2}){o_i}(t) + {\mu _1}{o_j}(t) + {\mu _2}{o_k}(t)\\
{o_j}^{ik}(t) = (1 - {\lambda _1} - {\lambda _2}){o_j}(t) + {\lambda _1}{o_i}(t) + {\lambda _2}{o_k}(t)\\
{o_k}^{ij}(t) = (1 - {\eta _1} - {\eta _2}){o_k}(t) + {\eta _1}{o_i}(t) + {\eta _2}{o_j}(t)
\end{array} \right.
\end{equation}

Otherwise:
\begin{equation}
    \left\{ \begin{array}{c}
{o_i}^{jk}(t) = {o_i}(t)\\
{o_j}^{ik}(t) = {o_j}(t)\\
{o_k}^{ij}(t) = {o_k}(t)
\end{array} \right.
\end{equation}
Where the values of  ${\mu _1} + {\mu _2}$, ${\lambda _1} + {\lambda _2}$ and ${\eta _1} + {\eta _2}$ are between $[0,0.5]$, and since the neighbors $j$ and $k$ of agent $i$ are not necessarily connected, the coefficient between them should be appropriately small, that is, the values of ${\lambda _2}$ and ${\eta _2}$ are between  $[0,0.2]$.

Then, after a time step, the agent's opinion value is updated as:
\begin{equation}
o_i^{}(t + 1) = o_i^{{t_n}}(t)
\end{equation}

Where ${t_n}$ represents the last point of opinion interaction within the time step  $t$.

(2)	Opinion natural evolution coefficient: According to\cite{30}, in real life, even if there is no interaction between agents, agent’s opinions will change over time, so we introduce the natural evolution coefficient of opinions into the interaction model. It is not difficult for us to understand that the agent's psychological cognitive ability is enhanced with the increase of knowledge and experience, so the natural evolution coefficient is not a fixed parameter, but a function that gradually increases over time, so it is a monotonically increasing function. In the early stage of opinion interaction, the value of this parameter should be small due to the limited knowledge of agents. As time goes by and agent’s knowledge and experience continue to increase, agent’s opinion choices may change, and the value of this parameter will become large. The value of this parameter should be in the range of  , therefore, we choose the derivative form of sigmoid function and add some coefficients to form the natural evolution coefficient of opinion as follows:

\begin{equation}
\alpha (t) = \frac{{\theta {e^{1/\beta (t - \gamma )}}}}{{{{(1 + {e^{1/\beta (t - \gamma )}})}^2}}}
\end{equation}

Where $\theta$  represents the influence index of $\alpha (t)$, which is generally 4.  $\beta $ represents the rate of natural evolution of opinions, which is between  $[1,4]$;  $\gamma $ represents the decay time of opinion, and  $\gamma $ can be taken as 1/10 of the total time step.

The natural evolution coefficient of opinions is used to modify the updating rules of opinions and the trust threshold function at the same time. The modified trust threshold function is expressed as follows:
\begin{equation}
    {\varepsilon _i}(t) = g(\left| {{o_i}(t)} \right|) = \alpha (t)({\varepsilon _{\min }} + \left| {{o_i}(t)} \right|({\varepsilon _{\max }} - {\varepsilon _{\min }}))
\end{equation}

After modification, the updating rules of an opinion interaction can be expressed as follows:
\begin{equation}
    \left\{ \begin{array}{c}
{o_i}^{jk}(t) = \alpha (t)[(1 - {\mu _1} - {\mu _2}){o_i}(t) + {\mu _1}{o_j}(t) + {\mu _2}{o_k}(t)]\\
{o_j}^{ik}(t) = \alpha (t)[(1 - {\lambda _1} - {\lambda _2}){o_j}(t) + {\lambda _1}{o_i}(t) + {\lambda _2}{o_k}(t)]\\
{o_k}^{ij}(t) = \alpha (t)[(1 - {\eta _1} - {\eta _2}){o_k}(t) + {\eta _1}{o_i}(t) + {\eta _2}{o_j}(t)]
\end{array} \right.
\end{equation}

Where the value of $\alpha (t)$ is only related to the time step and has nothing to do with the number of interactions between opinions, that is, within a time step, the value of  $\alpha (t)$ will not change even if multiple interactions occur.

\subsection{The interactive environment}
In the original D-W model, the interactive environment is relatively fixed and one-sided, and it is assumed that only one interaction of opinions is completed in each time step, which is not consistent with the interaction in reality. Therefore, in this paper, we consider that an agent can interact with all his neighbors within one step. In addition, selective disconnection mechanism and algorithm recommendation mechanism of social network are added in the interaction process to dynamically complete the information interaction between each agent.
\begin{enumerate}[(1)]
    \item Original social network: In the interaction analysis of agent opinions, we choose a variety of social network types for comparison, including scaling network, small-world network and random network. According to existing studies, it is found that the node degree of social networks in reality presents a power-law distribution. Therefore, when studying the influence of opinion leaders and active agents, we choose the scale-free network as our original network, so as to get closer to the reality.
    \item Selective edge reduction mechanism: Inspired by the node selection mechanism in \cite{31}, for randomly selected agent $i$, the trust threshold ${\varepsilon _i}(t + 1)$ will be calculated according to opinion ${o_i}(t + 1)$ after its opinion is updated, and selective disconnection will be carried out: For set  $A = \{ x \in N|\left| {{o_i}(t + 1) - {o_x}(t + 1)} \right| > {\varepsilon _i}(t + 1)\} $, agents $y \in A$ and $z \in A$ is randomly selected and the corresponding edges $(i,y) \in E$ and $(i,z) \in E$ of $i$ are deleted.
    \item Random edge addition mechanism: Also inspired by the node selection mechanism in \cite{31} for the randomly selected agent $i$, after selective disconnection, the neighbor will be re-selected according to the agents recommended by the algorithm: For set $B = \{ x \in {N_0}|\left| {{o_i}(t + 1) - {o_x}(t + 1)} \right| < 0.2,x \notin N\} $, The agent $y \in B$ and $y \in B$ was randomly selected and the corresponding edge of agent $i$ and agent $y$, $z$ was added to obtain the social network $G(V(t + 1),E(t + 1))$.
\end{enumerate}

In this paper,through selective disconnection mechanism and algorithm recommendation mechanism, the topology of social network can be dynamically changed to simulate the formation mechanism of echo chamber effect. With the passage of time and the interaction of opinions, agent’s information channels become increasingly narrow, and echo chambers are formed around different opinions and their own opinions are constantly strengthened in the echo chambers.

\section{Simulations}
\subsection{Data preprocessing}
According to the opinion interaction model set in the previous section, the influence of network structure, opinion leaders and active nodes on echo chamber effect is analyzed experimentally in this section.

Suppose there is a group of size $N=150$, and its agent composition is  ${N_0} = \{ 1,2,3, \cdots ,i, \cdots ,N\} $, which constitutes a social network $G(V,E)$. For the agent $i$, the original opinion value is set as any real number in the interval  $[-1,1]$,that is, ${o_i}(0) \in [ - 1,1]$ ,its opinion value at time $t$ is ${o_i}(t) \in [ - 1,1]$.

In the experiment, agents were divided into clusters according to opinion values  $o > 0.33$, $o < 0.33$, $ - 0.33 \le o \le 0.33$.  $A = \{ i|{o_i} > 0.33\}$ was a positive opinion cluster,  $B = \{ i|{o_i} <  - 0.33\} $ was a negative opinion cluster, and  $C = \{ i|\left| {{o_i}} \right| \le 0.33\} $ was a relatively neutral opinion cluster. In addition, according to the definition of average path length in complex networks \cite{32}, index   $k = ({c_1} + {c_2} + {c_3})/\frac{1}{2}T(T - 1)$ is further defined to measure the changes of individual information channels in the evolution of the model, where ${c_1}$,  ${c_2}$ and  ${c_3}$ respectively represent the number of connected edges among three different clusters, and  $T$ represents the total number of edges in the whole network.

\subsection{Echo chamber effect influenced by network structure}
In the study of social network, commonly used network models include scale-free network, small-world network, ER random networks and so on. In the above-mentioned social network model, the node degree distribution of scale-free network\cite{33} follows the power-law distribution. Most ordinary nodes have few connections, while a few authoritative nodes have many connections, which is consistent with the structure of social networks in reality, and therefore is often studied as the original network. For small-world network\cite{34}, the average path between nodes is much smaller than the number of nodes in the network, and has the properties of high aggregation and short average minimum path, which is often used to study the dynamic phenomena in social networks. In addition, ER random network\cite{35} usually has relatively average node degree and short average path, so it is often used as the reference object of the original network or other networks. Here, random network is used as the reference of scale-free network and small-world network. As be known, the formation of echo chamber effect is accompanied by the dynamic changes of social network topology. Therefore, studying the evolution process of social network opinions is beneficial to further understand its formation mechanism. Next, we use scale-free network, small-world network and ER random network respectively to study the formation mechanism of echo chamber effect. Specific experimental parameters and results are shown in the figure 1,2,3 and table 1.

\begin{center}
\textbf{Table1 Original network parameters}
\label{Table1}
\resizebox{\textwidth}{15mm}{
\begin{tabular}{ccccc}
   \toprule
   Network type & Number of nodes & average degree & Cluster coefficient & Average shortest path length \\
   \midrule
    random network & 150 & 8 & 0.048 & 2.616 \\
   Scale-free network & 150 & 7.78 & 0.130 & 2.528 \\
   Small-World network & 150 & 8 & 0.230 & 2.853 \\
   \bottomrule
\end{tabular}}
\end{center}

\begin{figure}[htbp]
	\centering
	\begin{minipage}{0.32\linewidth}
		\centering
		\includegraphics[width=0.9\linewidth]{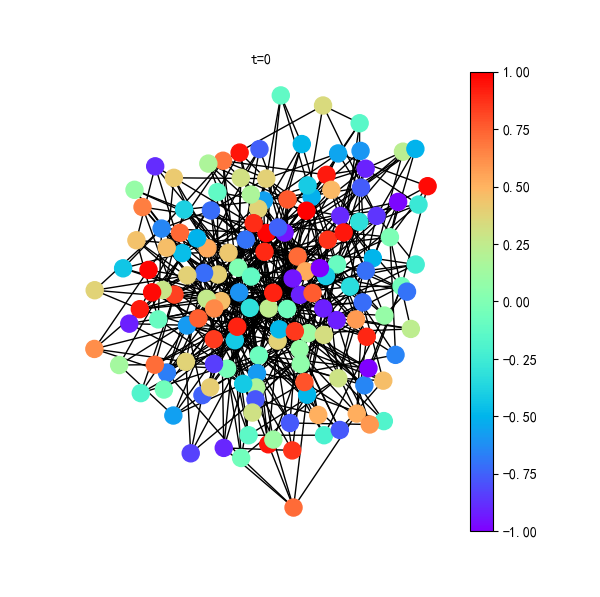}
		\subcaption{}
		\label{1-1}
	\end{minipage}
	\begin{minipage}{0.32\linewidth}
		\centering
		\includegraphics[width=0.9\linewidth]{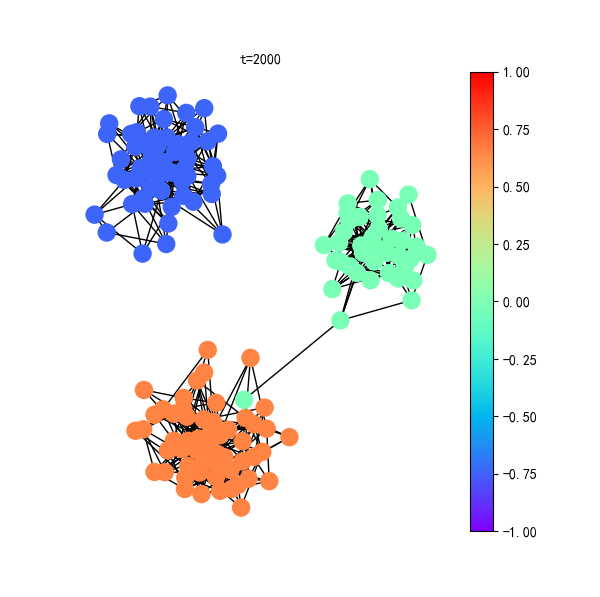}
		\subcaption{}
		\label{1-2}
	\end{minipage}
	\begin{minipage}{0.32\linewidth}
		\centering
		\includegraphics[width=0.9\linewidth]{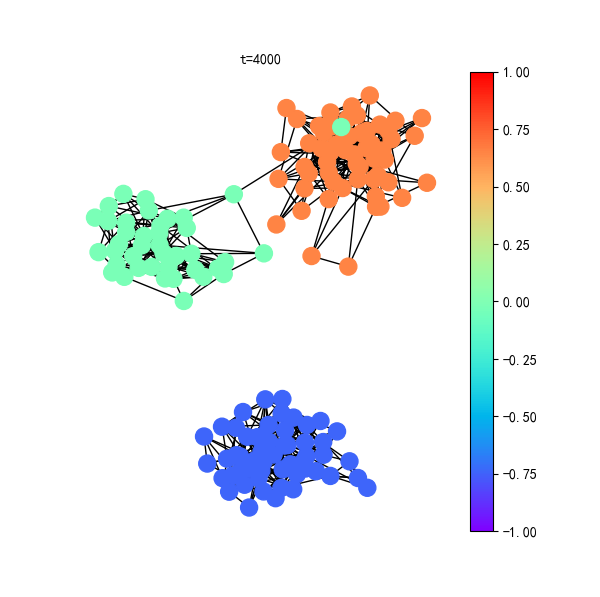}
		\subcaption{}
		\label{1-3}
	\end{minipage}
	\begin{minipage}{0.32\linewidth}
		\centering
		\includegraphics[width=0.9\linewidth]{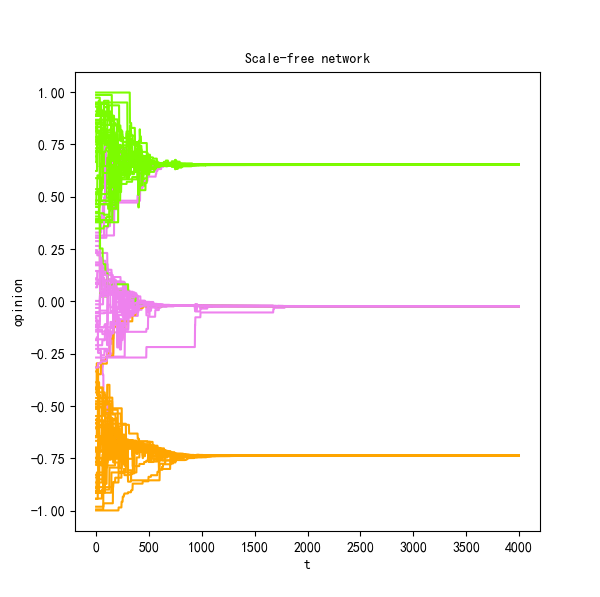}
		\subcaption{}
		\label{1-4}
	\end{minipage}
	\begin{minipage}{0.32\linewidth}
		\centering
		\includegraphics[width=0.9\linewidth]{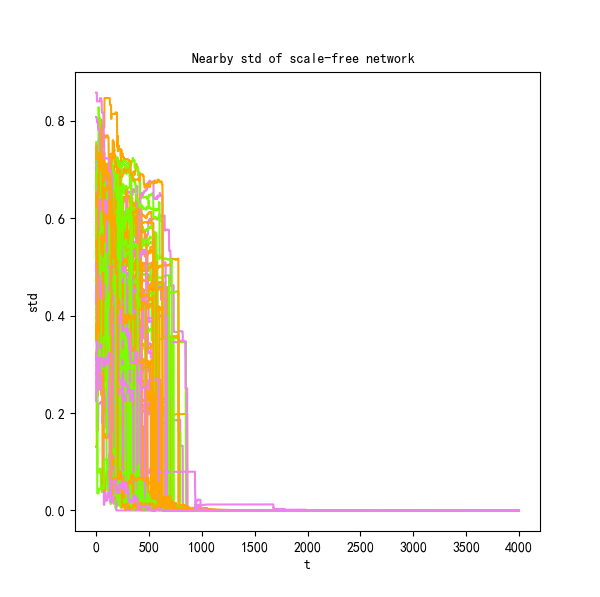}
		\subcaption{}
		\label{1-5}
	\end{minipage}
	\begin{minipage}{0.32\linewidth}
		\centering
		\includegraphics[width=0.9\linewidth]{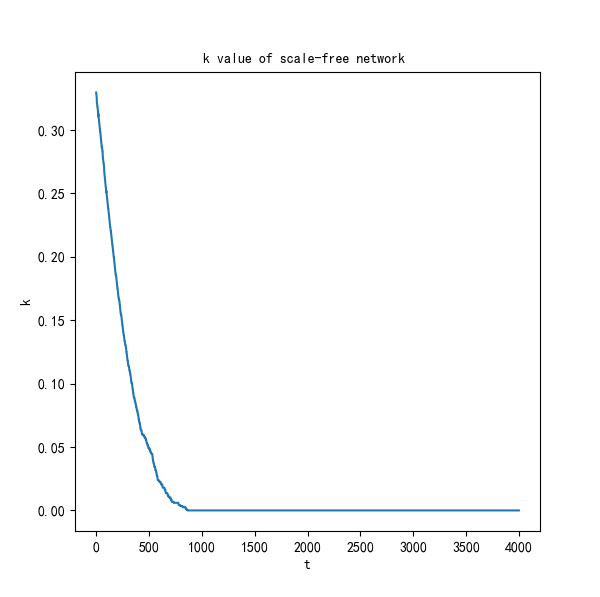}
		\subcaption{}
		\label{1-6}
	\end{minipage}
	\caption{Scale-free network opinion evolution diagram}
	\label{fig1}
\end{figure}

\begin{figure}[htbp]
	\centering
	\begin{minipage}{0.32\linewidth}
		\centering
		\includegraphics[width=0.9\linewidth]{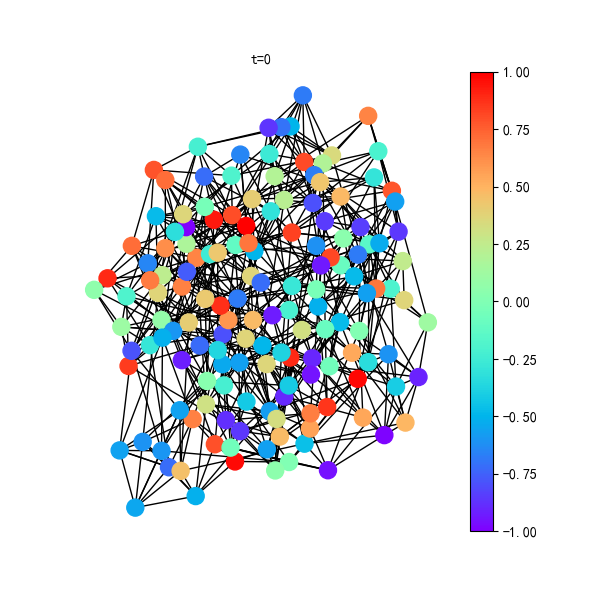}
		\subcaption{}
		\label{2-1}
	\end{minipage}
	\begin{minipage}{0.32\linewidth}
		\centering
		\includegraphics[width=0.9\linewidth]{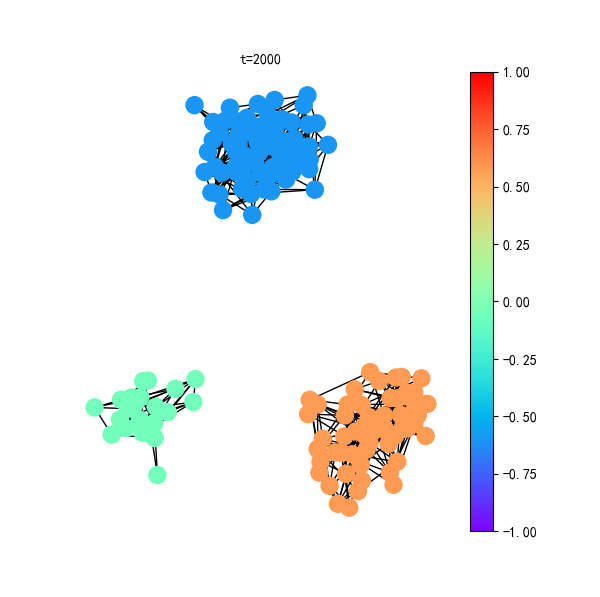}
		\subcaption{}
		\label{2-2}
	\end{minipage}
	\begin{minipage}{0.32\linewidth}
		\centering
		\includegraphics[width=0.9\linewidth]{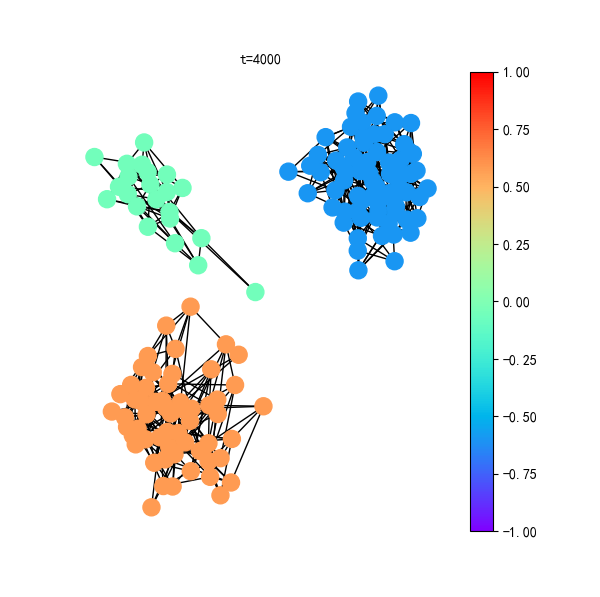}
		\subcaption{}
		\label{2-3}
	\end{minipage}
	\begin{minipage}{0.32\linewidth}
		\centering
		\includegraphics[width=0.9\linewidth]{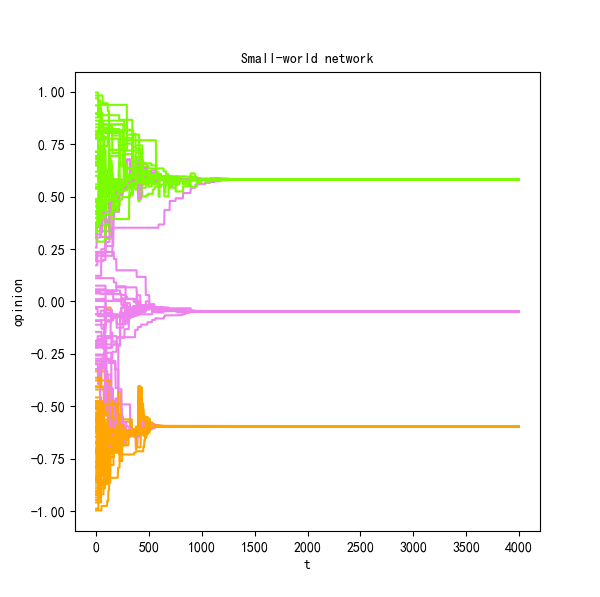}
		\subcaption{}
		\label{2-4}
	\end{minipage}
	\begin{minipage}{0.32\linewidth}
		\centering
		\includegraphics[width=0.9\linewidth]{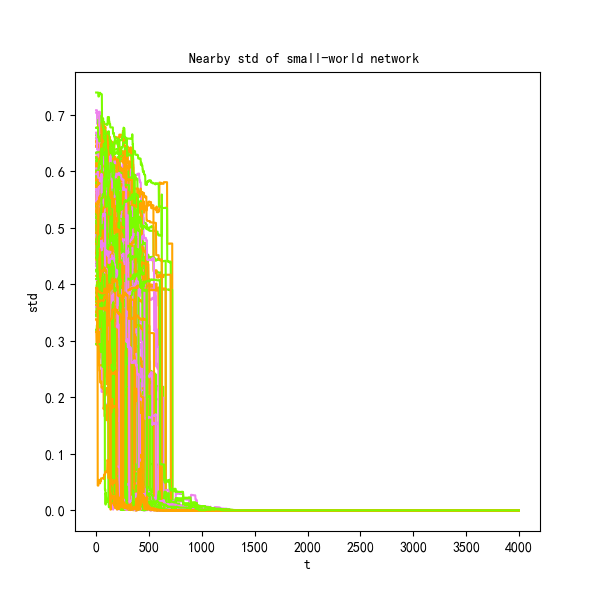}
		\subcaption{}
		\label{2-5}
	\end{minipage}
	\begin{minipage}{0.32\linewidth}
		\centering
		\includegraphics[width=0.9\linewidth]{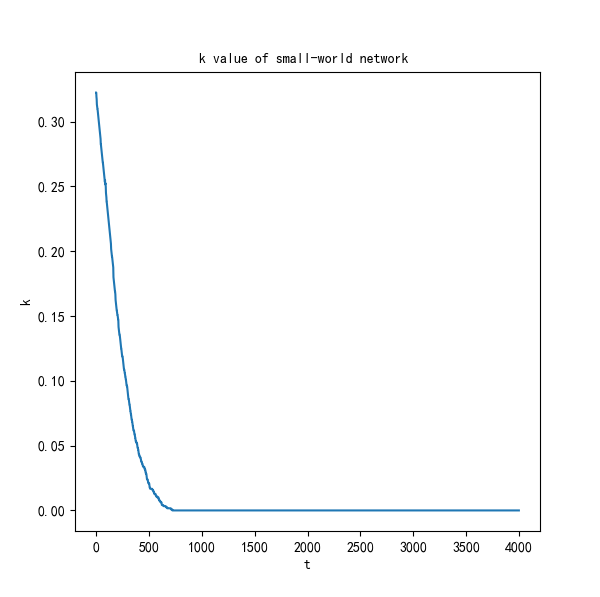}
		\subcaption{}
		\label{2-6}
	\end{minipage}
	\caption{Small-world network opinion evolution diagram}
	\label{fig2}
\end{figure}

\begin{figure}[htbp]
	\centering
	\begin{minipage}{0.32\linewidth}
		\centering
		\includegraphics[width=0.9\linewidth]{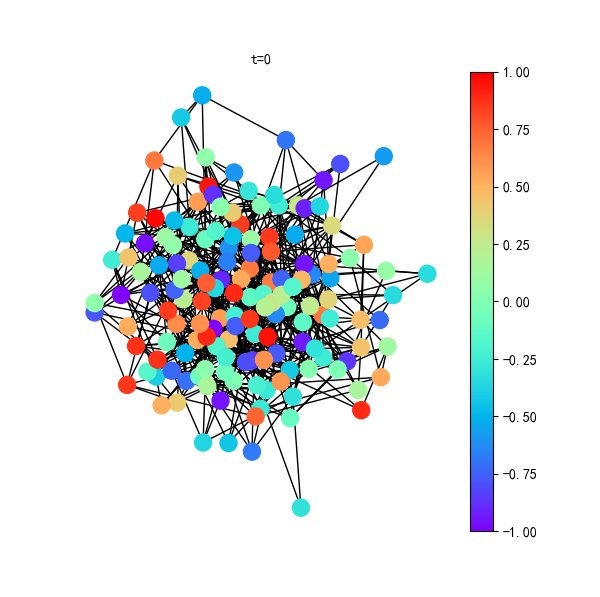}
		\subcaption{}
		\label{3-1}
	\end{minipage}
	\begin{minipage}{0.32\linewidth}
		\centering
		\includegraphics[width=0.9\linewidth]{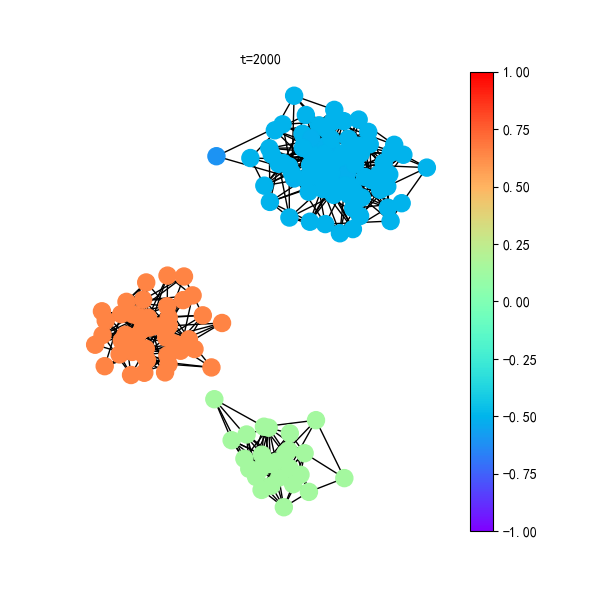}
		\subcaption{}
		\label{3-2}
	\end{minipage}
	\begin{minipage}{0.32\linewidth}
		\centering
		\includegraphics[width=0.9\linewidth]{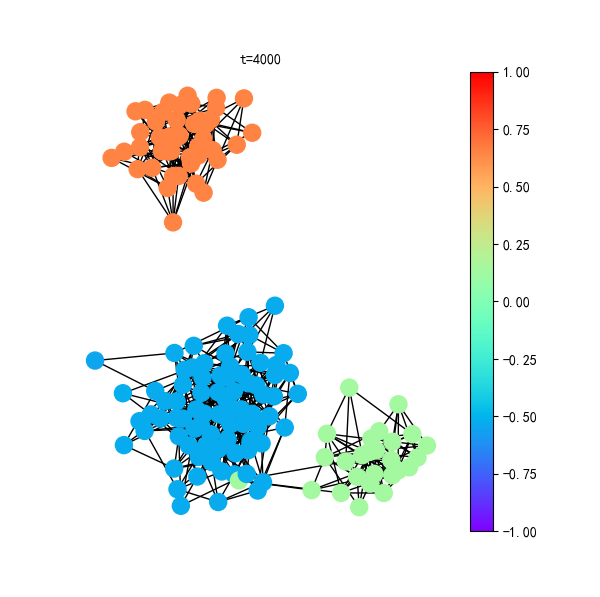}
		\subcaption{}
		\label{3-3}
	\end{minipage}
	\begin{minipage}{0.32\linewidth}
		\centering
		\includegraphics[width=0.9\linewidth]{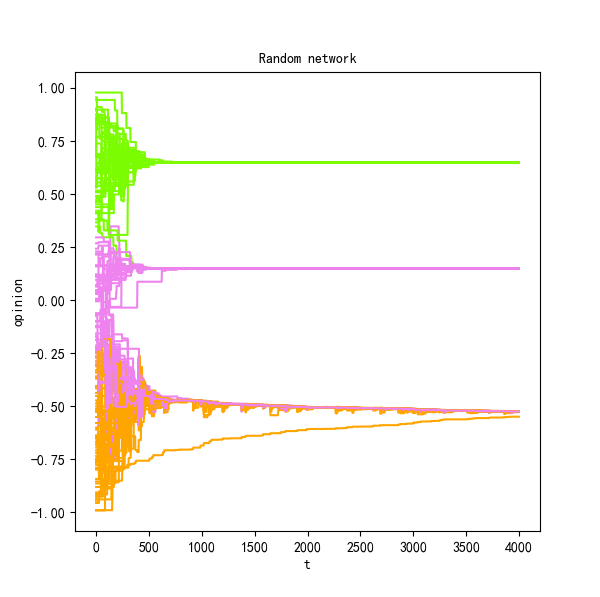}
		\subcaption{}
		\label{3-4}
	\end{minipage}
	\begin{minipage}{0.32\linewidth}
		\centering
		\includegraphics[width=0.9\linewidth]{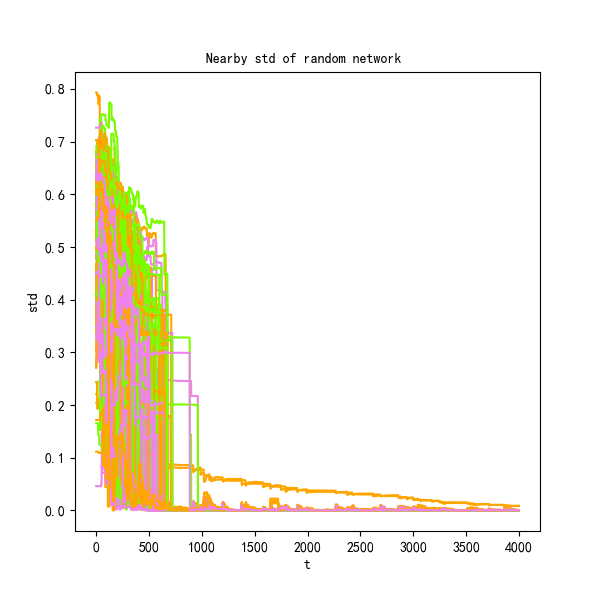}
		\subcaption{}
		\label{3-5}
	\end{minipage}
	\begin{minipage}{0.32\linewidth}
		\centering
		\includegraphics[width=0.9\linewidth]{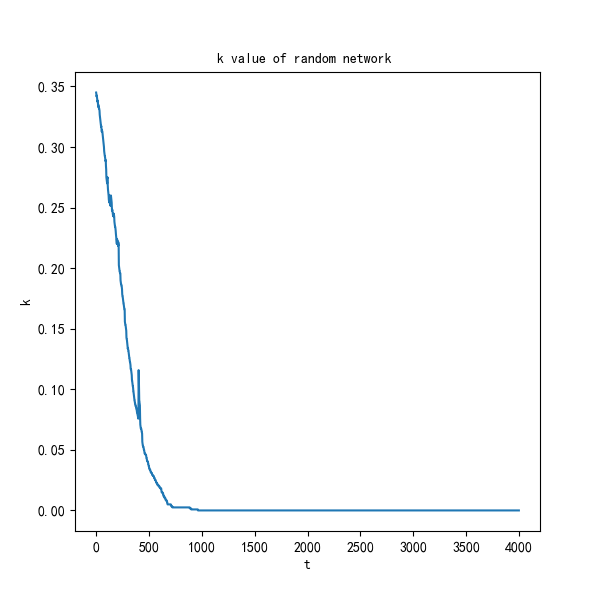}
		\subcaption{}
		\label{3-6}
	\end{minipage}
	\caption{Random network opinion evolution diagram}
	\label{fig3}
\end{figure}
According to the network evolution results in Figure 1, Figure 2 and Figure 3 above, biased aggregation of opinions and isolation of heterogeneous groups occurred in all three network structures, and information was transmitted only among homogeneous groups. In 4000 time step iterations, the networks of the three structures all form corresponding echo chambers at the opinion values  $o > 0.33$,$o < 0.33$, $ - 0.33 \le o \le 0.33$, and opinion is strengthened continuously, and finally a unified opinion is formed in the echo chamber. On the other hand, the neighbor standard deviation evolution diagram and k-value diagram under the three network structures describe the narrowing process of information channels between nodes and groups during the evolution of echo chamber effect model from the agent’s level and the group level respectively. It is not difficult to find that, under the three network structures, the neighbor standard deviation and K value decay rapidly, and the group is isolated in a short time, the opinions tend to be homogenous. Further, we consider the opinion aggregation process in the three networks. When $t \approx 1000$, the opinion aggregation in each region is completed, and when $t \approx 500$, its k value is roughly reduced to 0, that is, they complete the opinion aggregation under the condition of highly isolated heterogeneous groups.

From the perspective of network stability, we believe that the network with higher average node degree has better connectivity and stability, while the three networks we set have low average node degree, so the network has weak anti-interference. The echo chamber formation mechanism can change the topology of the network in a short time, and affect the aggregation process of opinions while separating heterogeneous groups. Therefore, it is considered to increase the average node degree and aggregation coefficient of the scale-free network, improve the stability and connectivity of the network, and explore the evolution of its opinions and the formation process of echo chamber effect.

\begin{figure}[htbp]
	\centering
	\begin{minipage}{0.49\linewidth}
		\centering
		\includegraphics[width=0.9\linewidth]{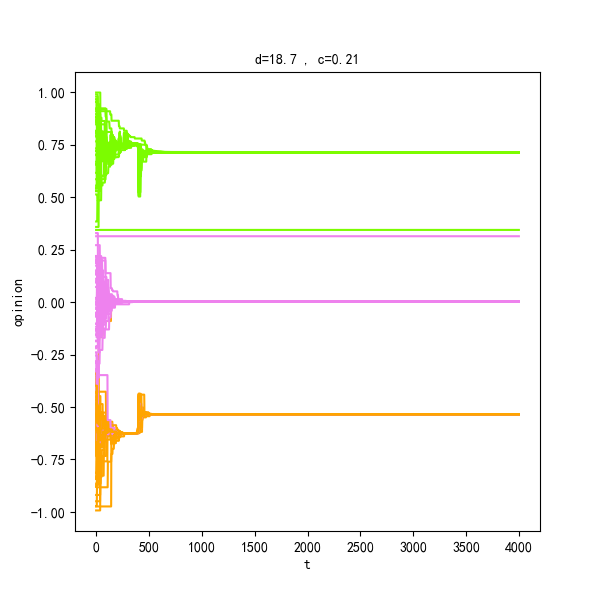}
		\subcaption{}
		\label{4-1}
	\end{minipage}
	\begin{minipage}{0.49\linewidth}
		\centering
		\includegraphics[width=0.9\linewidth]{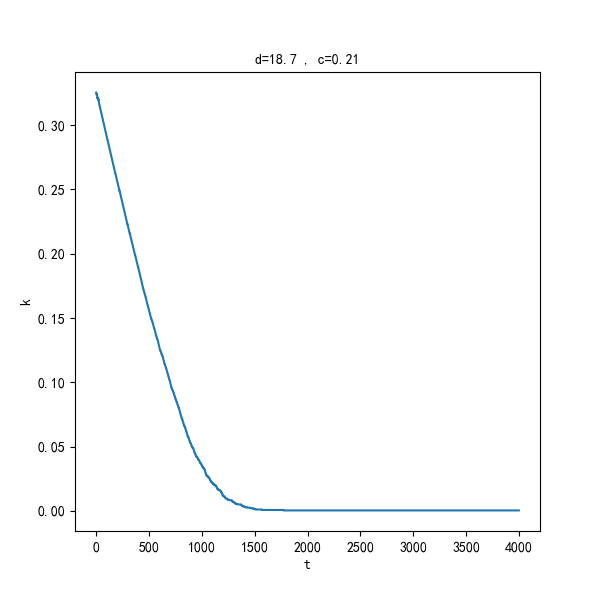}
		\subcaption{}
		\label{4-2}
	\end{minipage}
	
	\begin{minipage}{0.49\linewidth}
		\centering
		\includegraphics[width=0.9\linewidth]{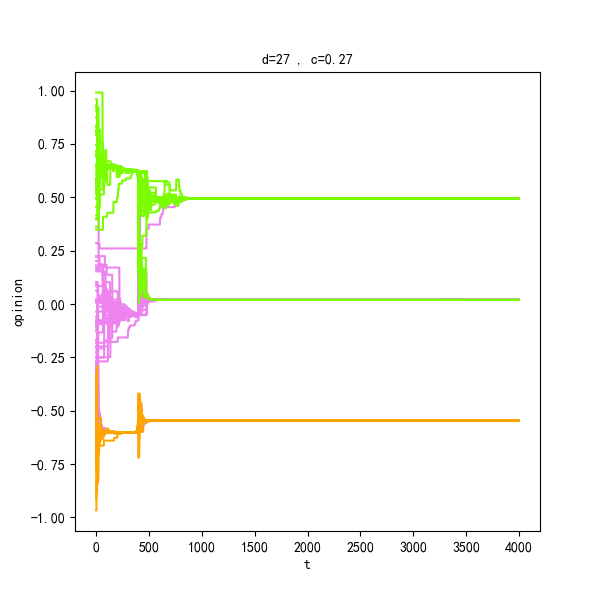}
		\subcaption{}
		\label{4-3}
	\end{minipage}
	\begin{minipage}{0.49\linewidth}
		\centering
		\includegraphics[width=0.9\linewidth]{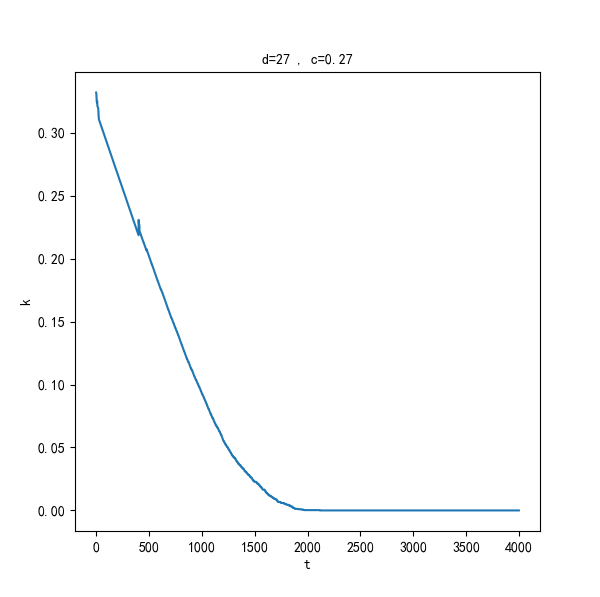}
		\subcaption{}
		\label{4-4}
	\end{minipage}
	\caption{Opinion evolution diagram after increasing node degree}
	\label{fig4}
\end{figure}

Figure 4 is the opinion evolution diagram and k-value curve after adding the mean node degree and aggregation coefficient of the scale-free network. It is not difficult to find that with the increase of network average node degree and aggregation coefficient, its k value curve tends to be flat, and the improvement of network stability and connectivity slows down the separation of heterogeneous groups. On the other hand, the improvement of network connectivity and stability speeds up the aggregation of opinions, which has been completed before $t \approx 500$. With the increase of network stability and connectivity, the separation of heterogeneous groups slows down, the formation of echo chamber becomes slow, and the aggregation of agent’s opinions has been completed before the group isolation.

In the process of opinion evolution, echo chamber effect cannot be eliminated only by changing the original network type, and the formation mechanism of echo chamber effect can rapidly change the network topology. In addition, although increasing the connectivity and stability of the network can slow down the formation of echo chamber and increase the convergence speed of opinions, the echo chamber effect cannot be eliminated eventually. Therefore, opinion leaders and active nodes will be introduced into the model in the next part to explore whether the echo chamber effect can be eliminated.

\subsection{Echo chamber effect influenced by opinion leaders}
In the interaction process of opinions in reality, there are often some agents with high authority, whose opinions can guide and aggregate the opinions in a certain range around them. Such agents can be called opinion leaders. In this part, we will consider the influence of opinion leaders on the interaction process of opinions.

Firstly, we consider the influence range of opinion leaders' opinions. As can be seen from the previous description, if an agent holds an opinion value of $o$, its trust threshold is $\varepsilon (t) = \alpha (t)(0.2 + 0.2\left| o \right|)$, then the acceptance range of opinion can be expressed as follows:

\begin{equation}
    \left\{ \begin{array}{l}
A = [(1 - 0.2\alpha (t))o - 0.2\alpha (t),(1 + 0.2\alpha (t))o + 0.2\alpha (t)],{\rm{ }}o > 0,\\
A = [(1 + 0.2\alpha (t))o - 0.2\alpha (t),(1 - 0.2\alpha (t))o + 0.2\alpha (t)],{\rm{ }}o \le 0
\end{array} \right.
\end{equation}

Assuming that the opinion value of the opinion leader is , then, after calculation, its influence range can be expressed as the following formula:

\begin{equation}
    \left\{ \begin{array}{l}
{I_x} = \left[ {\frac{{x - 0.2\alpha (t)}}{{1 + 0.2\alpha (t)}},\frac{{x + 0.2\alpha (t)}}{{1 - 0.2\alpha (t)}}} \right],{\rm{ }}x \ge 0.2\alpha (t)\\
{I_x} = \left[ {\frac{{x - 0.2\alpha (t)}}{{1 - 0.2\alpha (t)}},\frac{{x + 0.2\alpha (t)}}{{1 + 0.2\alpha (t)}}} \right],{\rm{ }}x \le  - 0.2\alpha (t)\\
{I_x} = \left[ {\frac{{x - 0.2\alpha (t)}}{{1 - 0.2\alpha (t)}},\frac{{x + 0.2\alpha (t)}}{{1 - 0.2\alpha (t)}}} \right],{\rm{ }} - 0.2\alpha (t){\rm{ < }}x < 0.2\alpha (t)
\end{array} \right.
\end{equation}

According to the above formula, the influence range of opinion leaders is related to only the time step. After defining the influence range of opinion leaders, we still choose the scale-free network as our original network, and set the original average node degree of the scale-free network to 7.8. An opinion leader is introduced into the model to study the influence of opinion leaders on the formation of echo chamber. In the process of opinion interaction, in order to prevent opinion leader nodes from degenerating into ordinary nodes, we set the node degree of opinion leader to  $d \approx 149$, that is, opinion leader is connected with every other node. Now, we set the opinion values of opinion leaders as $o = 0,0.5,0.8$ respectively, and assume that opinion leader nodes will not interact with other common nodes, but can only be used as objects for reference, so as to ensure that opinion leaders' opinions will not change. In addition, we set parameter  ${k_l} = {c_1}/T$ according to the definition of aggregation coefficient in complex networks \cite{32}, and use ${k_l}$ to represent the evolution of information channels within the influence range of opinion leaders, ${c_1}$ to represent the number of connected edges between opinion leaders and points within the influence range, and  $T$ to represent the total number of all connected edges in the network.

When the time step $t=0$, the natural evolution coefficient of opinion is $\alpha (0) \approx 1$. Therefore, when the opinion value of opinion leaders is $o = 0,0.5,0.8$, their corresponding influence ranges are  ${I_{{x_1}}} = [ - 0.25,0.25],{I_{{x_2}}} = [0.25,0.875],{I_{{x_3}}} = [0.5,1]$ respectively. The corresponding experimental results are shown as follows:

\begin{figure}[htbp]
	\centering
	\begin{minipage}{0.32\linewidth}
		\centering
		\includegraphics[width=0.9\linewidth]{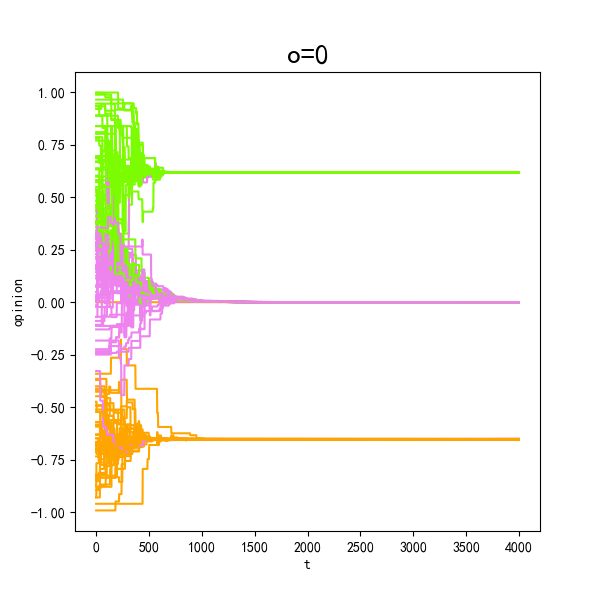}
		\subcaption{}
		\label{5-1}
	\end{minipage}
	\begin{minipage}{0.32\linewidth}
		\centering
		\includegraphics[width=0.9\linewidth]{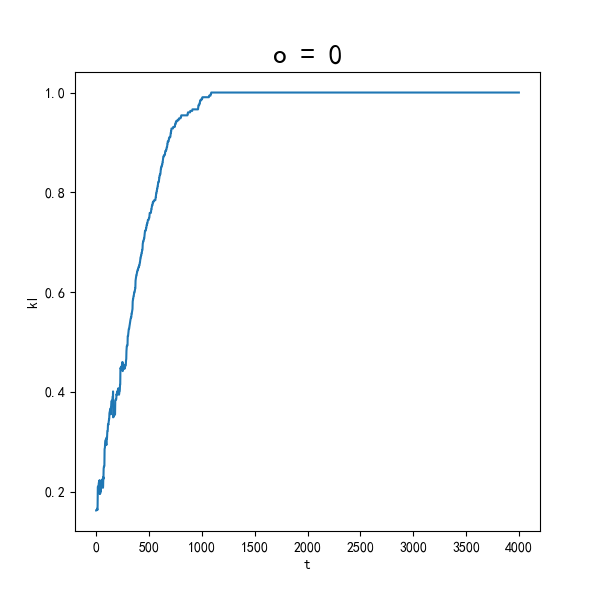}
		\subcaption{}
		\label{5-2}
	\end{minipage}
	\begin{minipage}{0.32\linewidth}
		\centering
		\includegraphics[width=0.9\linewidth]{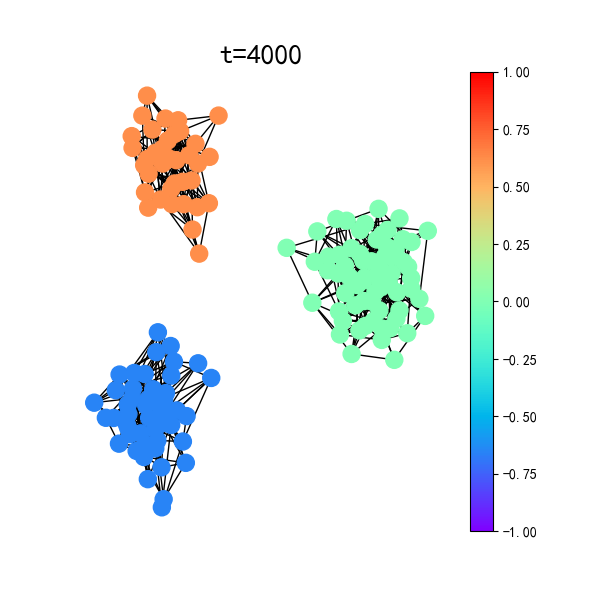}
		\subcaption{}
		\label{5-3}
	\end{minipage}
	\begin{minipage}{0.32\linewidth}
		\centering
		\includegraphics[width=0.9\linewidth]{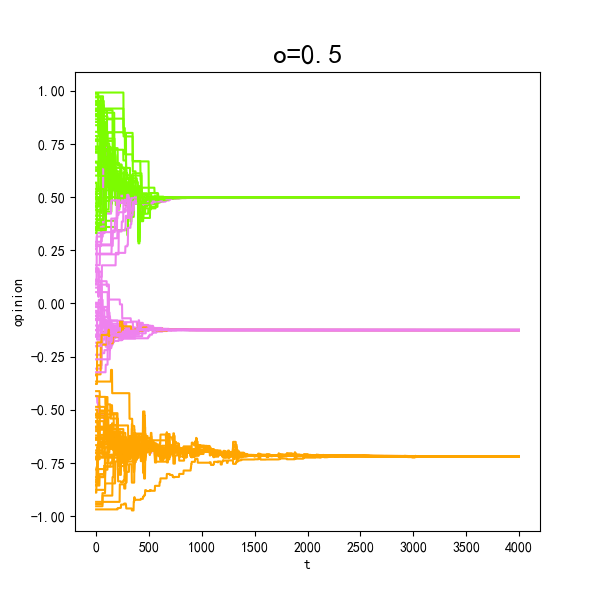}
		\subcaption{}
		\label{5-4}
	\end{minipage}
	\begin{minipage}{0.32\linewidth}
		\centering
		\includegraphics[width=0.9\linewidth]{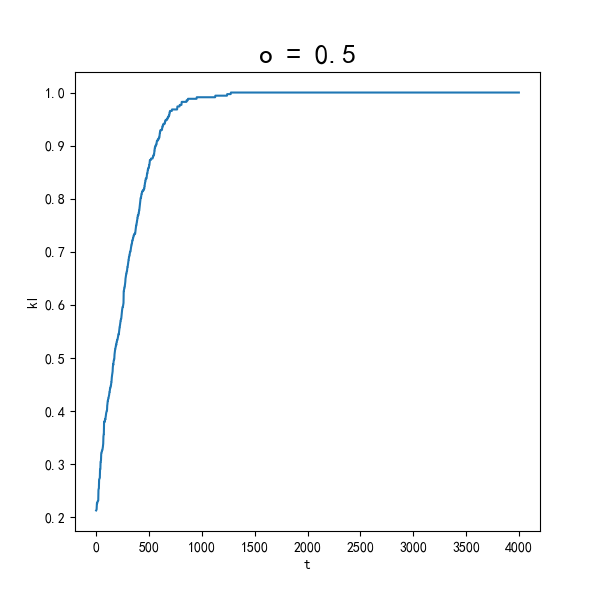}
		\subcaption{}
		\label{5-5}
	\end{minipage}
	\begin{minipage}{0.32\linewidth}
		\centering
		\includegraphics[width=0.9\linewidth]{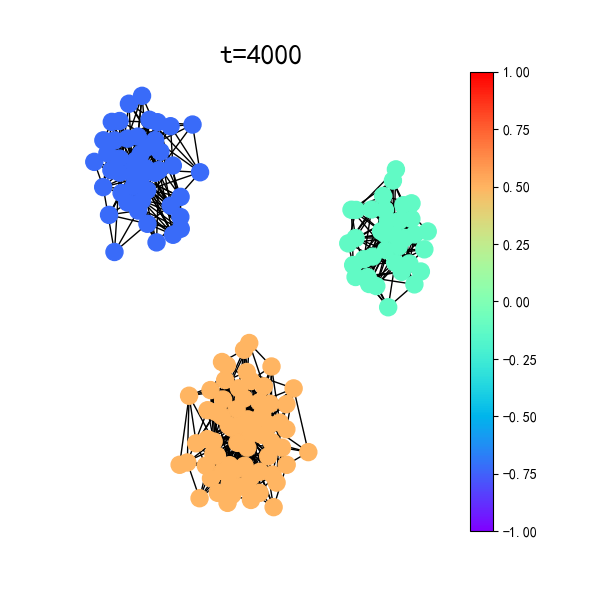}
		\subcaption{}
		\label{5-6}
	\end{minipage}
	\begin{minipage}{0.32\linewidth}
		\centering
		\includegraphics[width=0.9\linewidth]{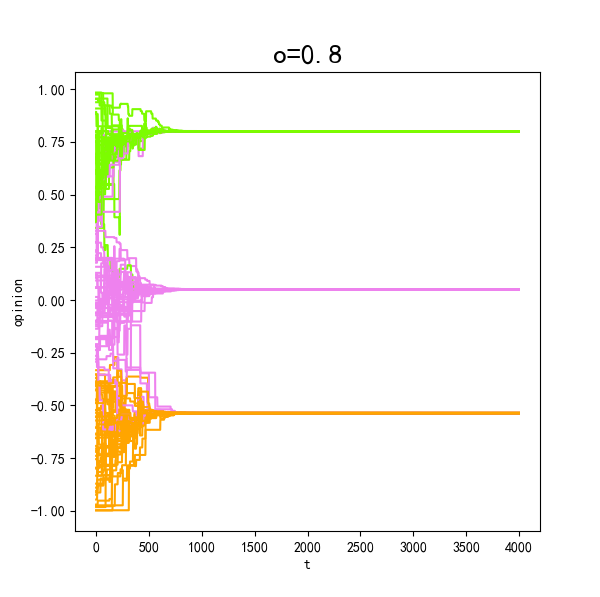}
		\subcaption{}
		\label{5-7}
	\end{minipage}
	\begin{minipage}{0.32\linewidth}
		\centering
		\includegraphics[width=0.9\linewidth]{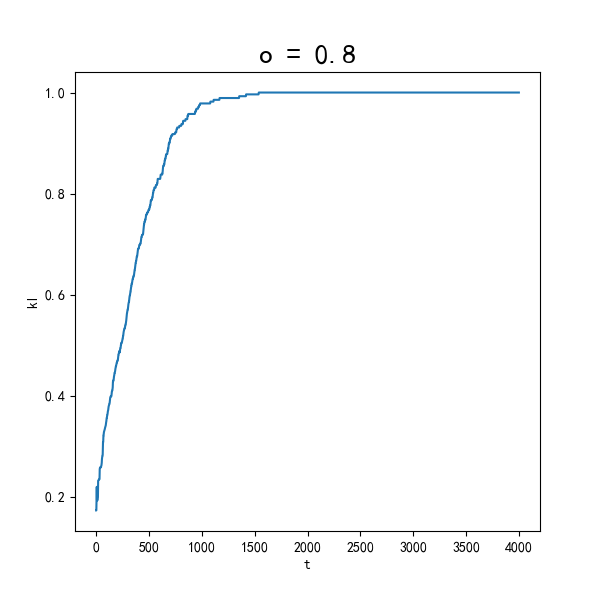}
		\subcaption{}
		\label{5-8}
	\end{minipage}
	\begin{minipage}{0.32\linewidth}
		\centering
		\includegraphics[width=0.9\linewidth]{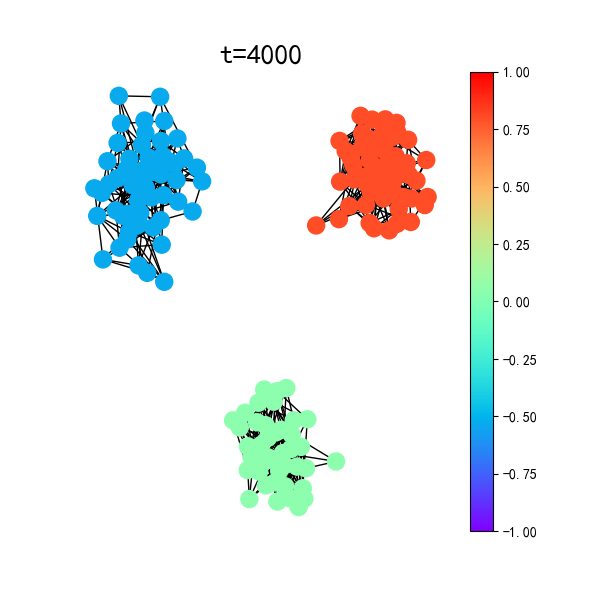}
		\subcaption{}
		\label{5-9}
	\end{minipage}
	\caption{Opinion evolution diagram under the influence of opinion leaders}
	\label{fig5}
\end{figure}

According to the results shown in figure 5, it is not difficult to find that when the opinion values of opinion leaders are set to 0,0.5 and 0.8 respectively, the opinion values of nodes within the influence range of opinion leaders will converge to the opinion values of opinion leaders, and corresponding opinion clusters will be formed near the opinion values of opinion leaders. In addition, we can also find from the K-value diagram that the information channels of nodes within the influence range of opinion leaders are rapidly narrowed and rapidly isolated from nodes outside the influence range. Their information mainly comes from opinion leaders. However, due to the limited influence range of opinion leaders, all opinions in the network cannot be aggregated together, so echo chambers are still formed.

In addition, Figure 6 shows the evolution process of nodes outside the influence range of opinion leaders. It can be found that no matter what the opinion value of opinion leaders is, some nodes outside the influence range will evolve dynamically to the vicinity of the opinion value of opinion leaders. Therefore, we believe that opinion leaders can aggregate opinions across the range.

\begin{figure}[htbp]
	\centering
	\begin{minipage}{0.49\linewidth}
		\centering
		\includegraphics[width=0.9\linewidth]{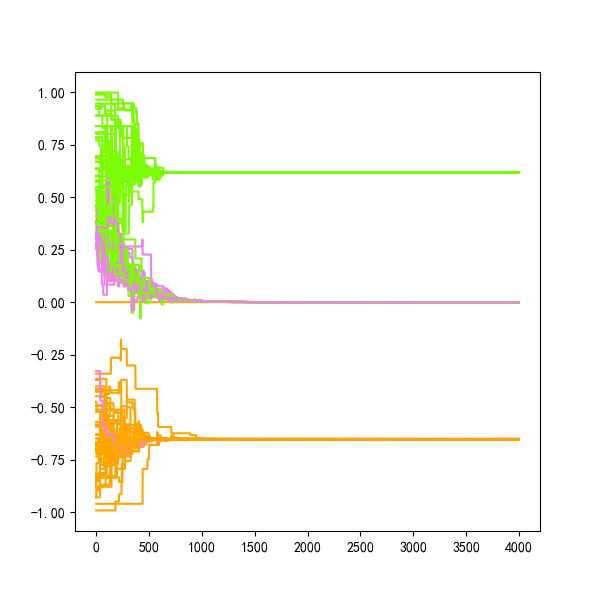}
		\subcaption{}
		\label{6-1}
	\end{minipage}
	\begin{minipage}{0.49\linewidth}
		\centering
		\includegraphics[width=0.9\linewidth]{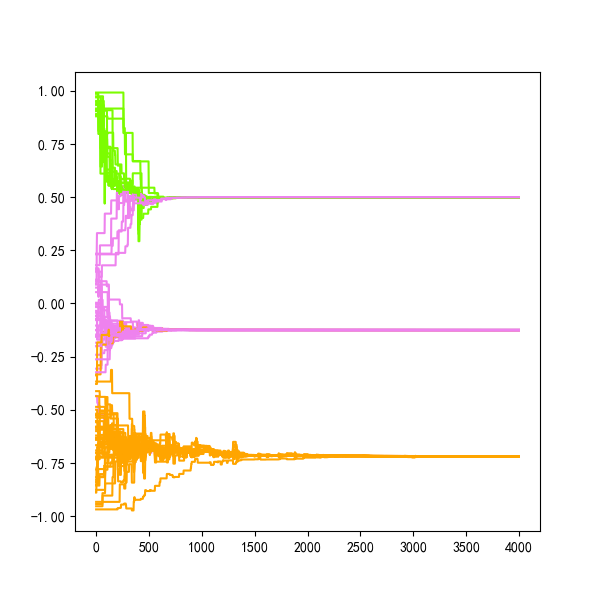}
		\subcaption{}
		\label{6-2}
	\end{minipage}
	\begin{minipage}{0.49\linewidth}
		\centering
		\includegraphics[width=0.9\linewidth]{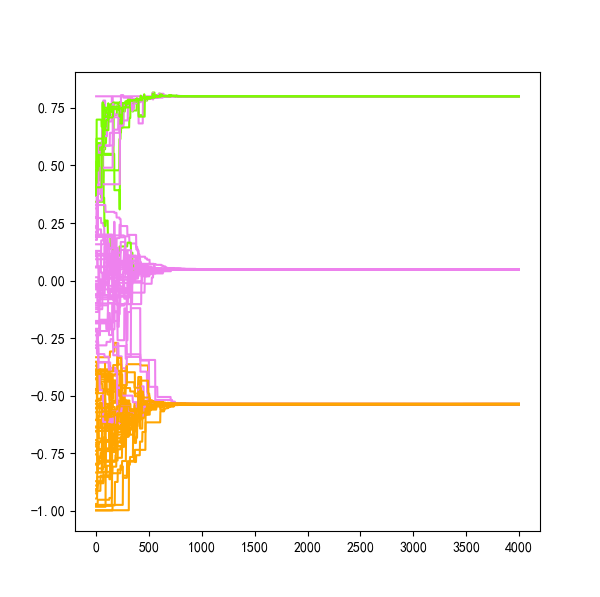}
		\subcaption{}
		\label{6-3}
	\end{minipage}
	\caption{Opinion evolution outside the influence of opinion leaders}
	\label{fig6}
\end{figure}

Considering the rapid aggregation of opinions within the influence range by opinion leaders, if we make the opinions of opinion leaders tend to be neutral over time, can the opinion value of us eventually converge near 0 through the guidance of opinion leaders? We add three opinion leaders into the model, and their original opinion values are 0.75,0 and -0.75 respectively. As time goes by, opinion leaders with opinion values of 0.75 and -0.75 will gradually approach to 0, while opinion leaders with opinion values of 0 remain unchanged. The results of opinion evolution guided by opinion leaders are shown as follows:

\begin{figure}[htbp]
	\centering
	\begin{minipage}{0.49\linewidth}
		\centering
		\includegraphics[width=0.9\linewidth]{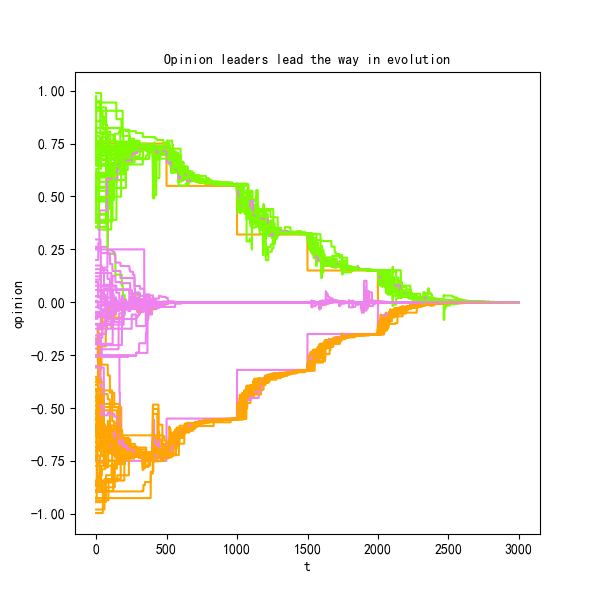}
		\subcaption{}
		\label{7-2}
	\end{minipage}
	\begin{minipage}{0.49\linewidth}
		\centering
		\includegraphics[width=0.9\linewidth]{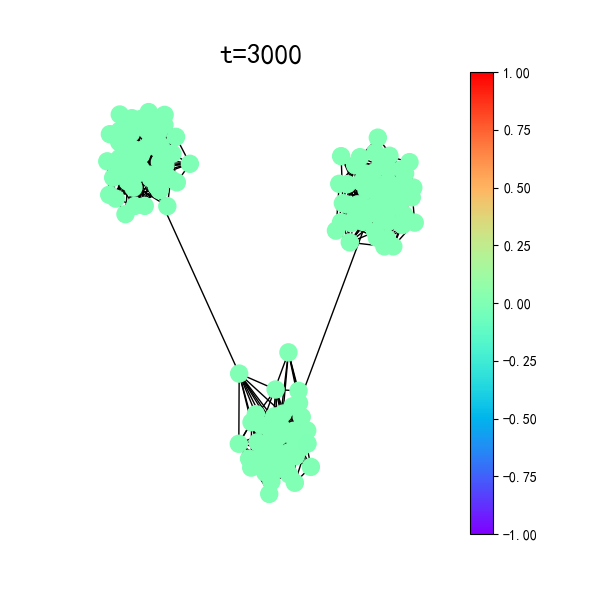}
		\subcaption{}
		\label{7-1}
	\end{minipage}
	\caption{Opinion evolution guided by opinion leaders}
	\label{fig7}
\end{figure}

It is not difficult to find from Figure 7 that when $t=500$, opinion values are aggregated around the three opinion leaders respectively. As the opinion values of opinion leaders approach 0 step by step, the opinion values near the opinion leaders also approach 0 along with the opinion leaders. Finally, when the opinions of almost all nodes converge to near 0, complete the aggregation of opinions, the group isolation phenomenon disappears, and the echo chamber ceases to exist.

From the perspective of model, opinion leaders can influence nodes within a certain range, so that their opinion values aggregate around the opinion values of opinion leaders, forming an echo chamber. However, it is not difficult to find that no matter what the opinion value of opinion leaders is, their scope of influence is limited. For some extremely stubborn nodes, opinion leaders cannot guide their opinions. In real life, if some authoritative media only express some neutral opinions, it can only aggregate the opinions of some people, and it cannot guide the opinions outside its influence range. Therefore, in order to break the echo chamber effect, the opinions of opinion leaders should not be unchanged,it is necessary to express opinions other than neutral ones to guide agent’s opinions.

\subsection{Echo chamber effect influenced by active nodes}
According to the above research, the information channel of low mean node degree network is rapidly narrowed, and heterogeneous groups are rapidly separated, thus forming one or more echo chambers. In order to slow down or even break the echo chamber effect, in this section, we will consider introducing active nodes into the model to increase the communication between heterogeneous groups, and investigate its impact on the formation of the echo chamber effect.

The interaction rules of active nodes are different from those of ordinary nodes. Active agents can actively contact heterogeneous information and tolerate neighbors with different opinions, which is specifically represented in the model as follows:

\begin{enumerate}[(1)]
    \item After the opinion update, nodes with heterogeneous opinions are randomly selected from all non-neighbor nodes $({o_a}{o_j} < 0)$, and their opinions are used as reference to update opinions again according to the given trust threshold $\varepsilon  = 1$.
    \item Does not change the interaction environment according to the selective disconnection mechanism. In addition, the fixed number of neighbors of active agents is set as $n=7$, and when $n>7$, a neighbor is randomly deleted. When $n<7$, the neighbor is added according to the algorithm recommendation mechanism.

\end{enumerate}

Here, we also set the scale-free network as the original network, and its original average node degree is also set to 7.8. To explore the influence of different number of active nodes in the model on the formation of echo chamber effect, we set the number of active nodes to 10, 30 and 50 respectively. The specific experimental results are shown as follows:

\begin{figure}[htbp]
	\centering
	\begin{minipage}{0.32\linewidth}
		\centering
		\includegraphics[width=0.9\linewidth]{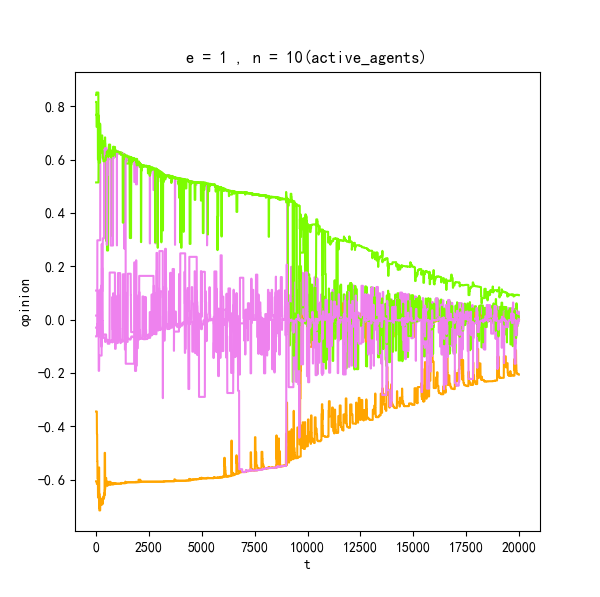}
		\subcaption{}
		\label{8-1}
	\end{minipage}
	\begin{minipage}{0.32\linewidth}
		\centering
		\includegraphics[width=0.9\linewidth]{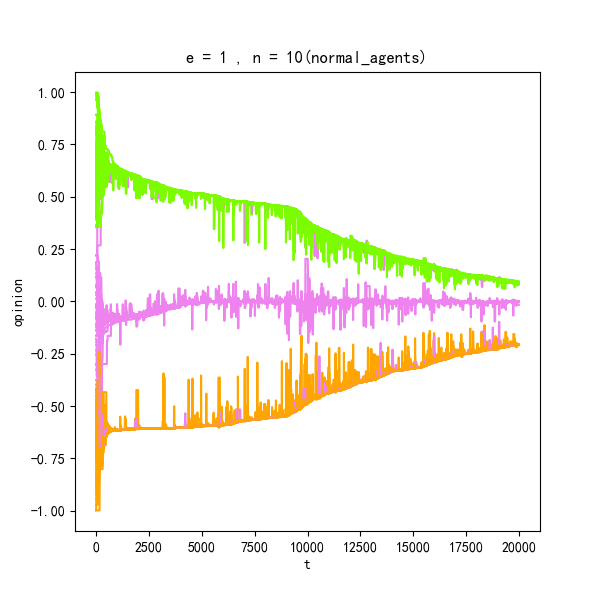}
		\subcaption{}
		\label{8-2}
	\end{minipage}
	\begin{minipage}{0.32\linewidth}
		\centering
		\includegraphics[width=0.9\linewidth]{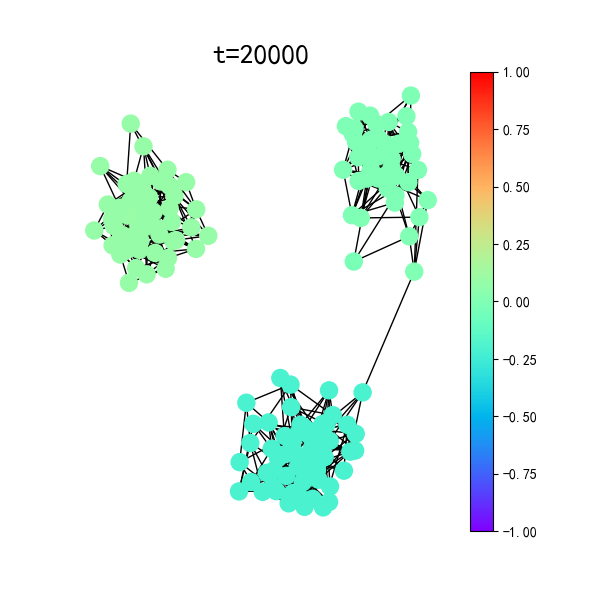}
		\subcaption{}
		\label{8-3}
	\end{minipage}
	\begin{minipage}{0.32\linewidth}
		\centering
		\includegraphics[width=0.9\linewidth]{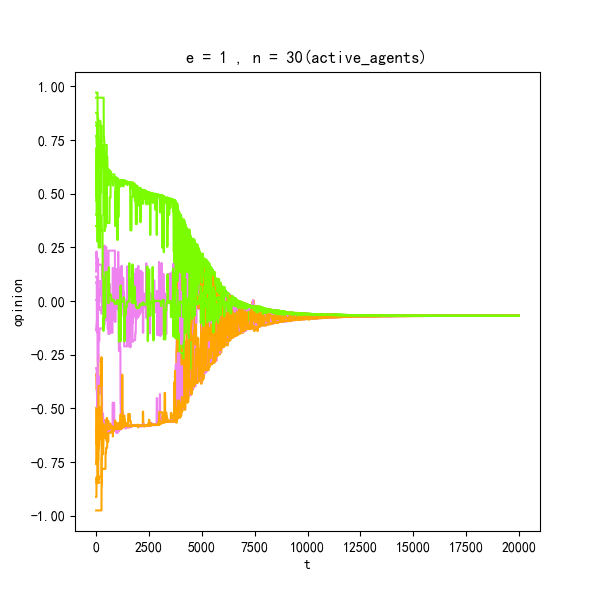}
		\subcaption{}
		\label{8-4}
	\end{minipage}
	\begin{minipage}{0.32\linewidth}
		\centering
		\includegraphics[width=0.9\linewidth]{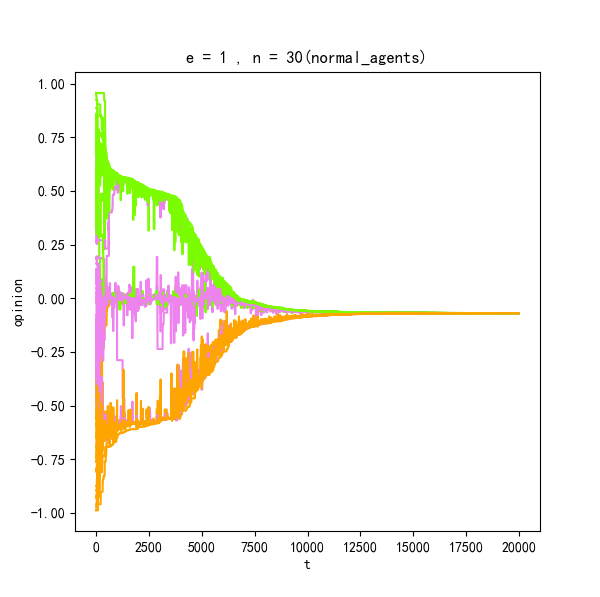}
		\subcaption{}
		\label{8-5}
	\end{minipage}
	\begin{minipage}{0.32\linewidth}
		\centering
		\includegraphics[width=0.9\linewidth]{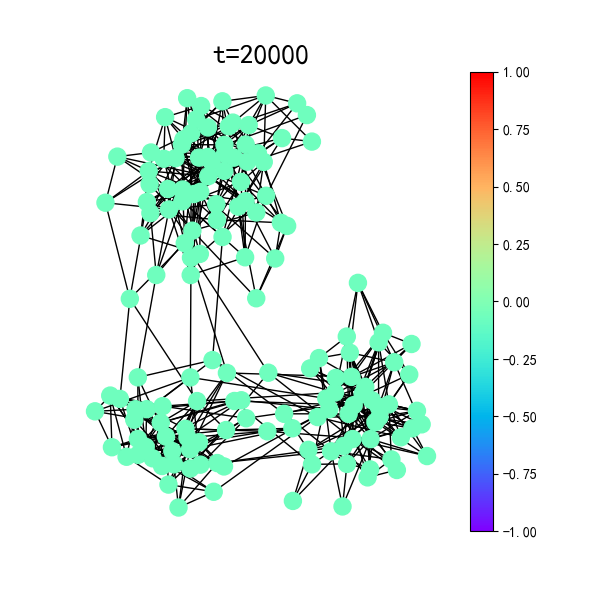}
		\subcaption{}
		\label{8-6}
	\end{minipage}
	\begin{minipage}{0.32\linewidth}
		\centering
		\includegraphics[width=0.9\linewidth]{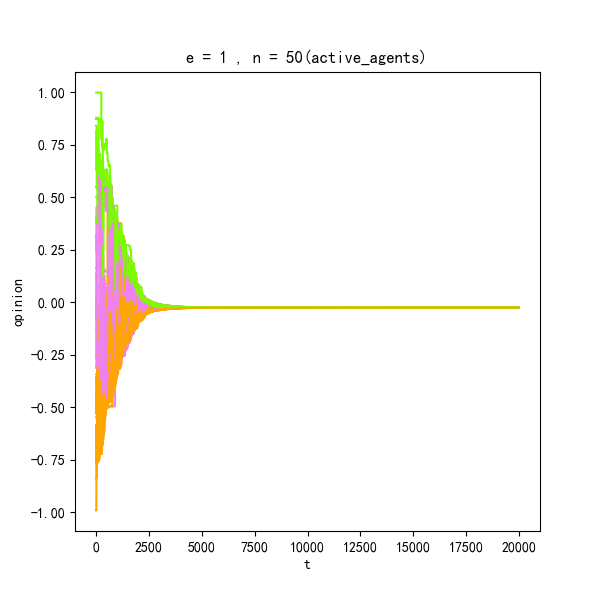}
		\subcaption{}
		\label{8-7}
	\end{minipage}
	\begin{minipage}{0.32\linewidth}
		\centering
		\includegraphics[width=0.9\linewidth]{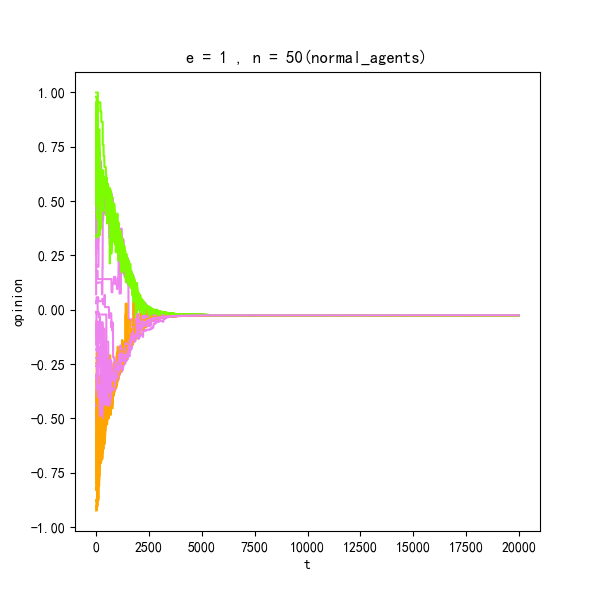}
		\subcaption{}
		\label{8-8}
	\end{minipage}
	\begin{minipage}{0.32\linewidth}
		\centering
		\includegraphics[width=0.9\linewidth]{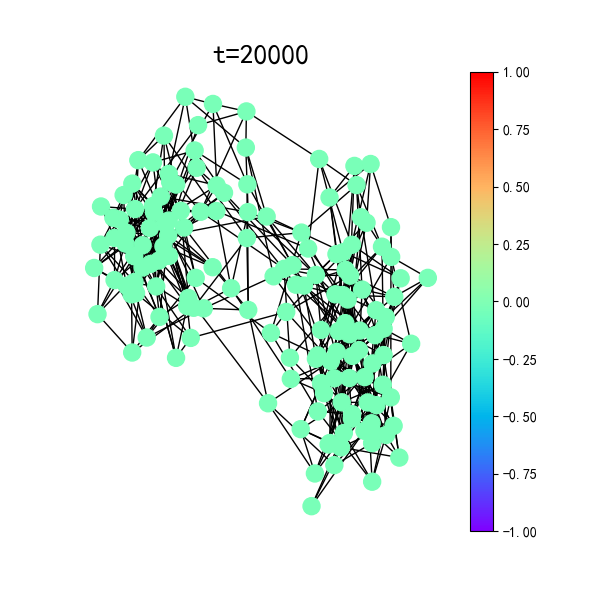}
		\subcaption{}
		\label{8-9}
	\end{minipage}
	\caption{The evolution of opinions under different numbers of active nodes}
	\label{fig8}
\end{figure}

As can be seen from Figure 8, after the introduction of active agents into the model, opinions will eventually converge to the vicinity of neutral opinions, and the convergence speed will accelerate with the increase of the number of active agents. Under different numbers of active agents, there will be no obvious phenomenon of heterogeneous group separation, that is, no echo chamber is formed. Therefore, it can be considered that, the addition of active agents effectively avoids the echo chamber effect.

In order to further explore the influence of active nodes on the formation of echo chamber effect, we specifically consider the evolution of opinions when the number of active nodes is 10, as shown in the figure 9:

\begin{figure}[htbp]
	\centering
	\begin{minipage}{0.32\linewidth}
		\centering
		\includegraphics[width=0.9\linewidth]{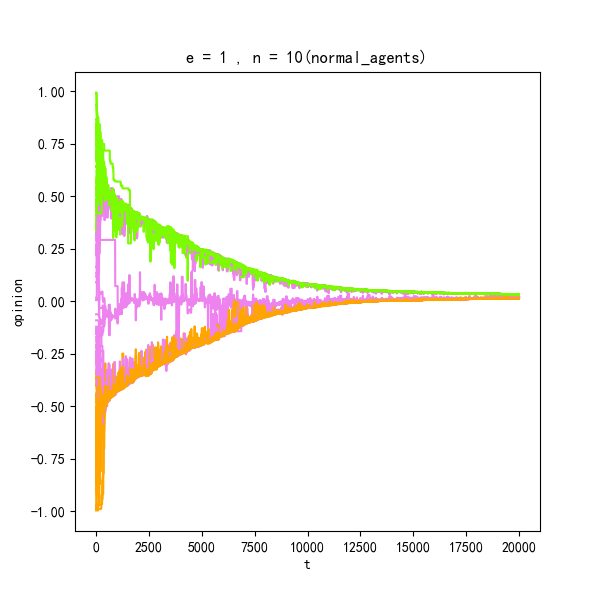}
		\subcaption{}
		\label{9-1}
	\end{minipage}
	\begin{minipage}{0.32\linewidth}
		\centering
		\includegraphics[width=0.9\linewidth]{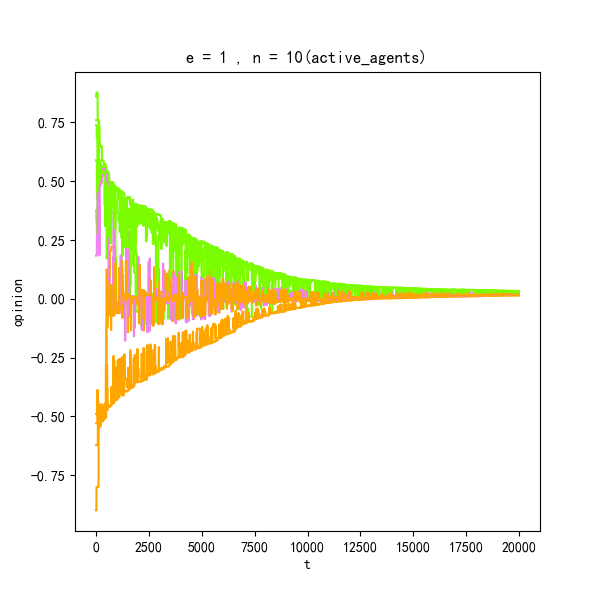}
		\subcaption{}
		\label{9-2}
	\end{minipage}
	\begin{minipage}{0.32\linewidth}
		\centering
		\includegraphics[width=0.9\linewidth]{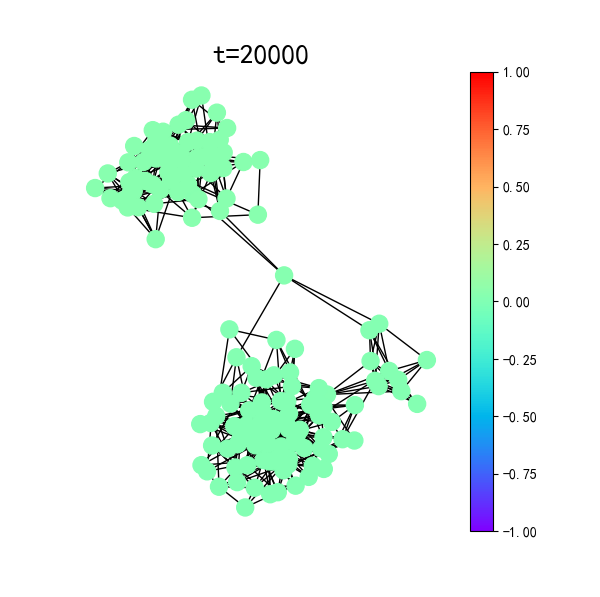}
		\subcaption{}
		\label{9-3}
	\end{minipage}
	\begin{minipage}{0.32\linewidth}
		\centering
		\includegraphics[width=0.9\linewidth]{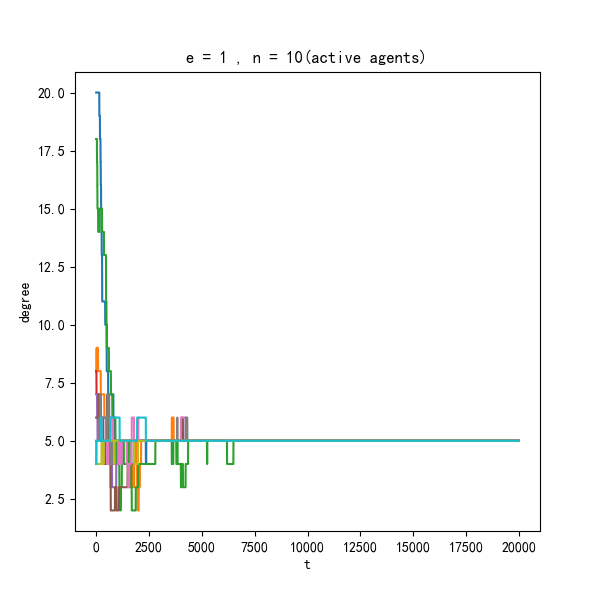}
		\subcaption{}
		\label{9-4}
	\end{minipage}
	\begin{minipage}{0.32\linewidth}
		\centering
		\includegraphics[width=0.9\linewidth]{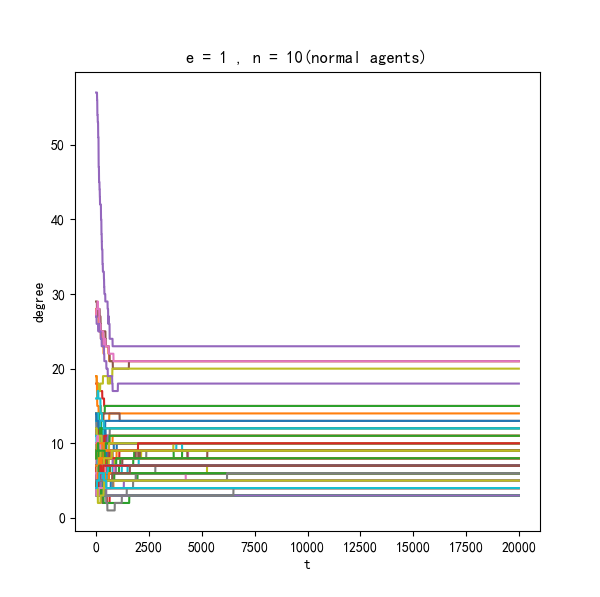}
		\subcaption{}
		\label{9-5}
	\end{minipage}
	\begin{minipage}{0.32\linewidth}
		\centering
		\includegraphics[width=0.9\linewidth]{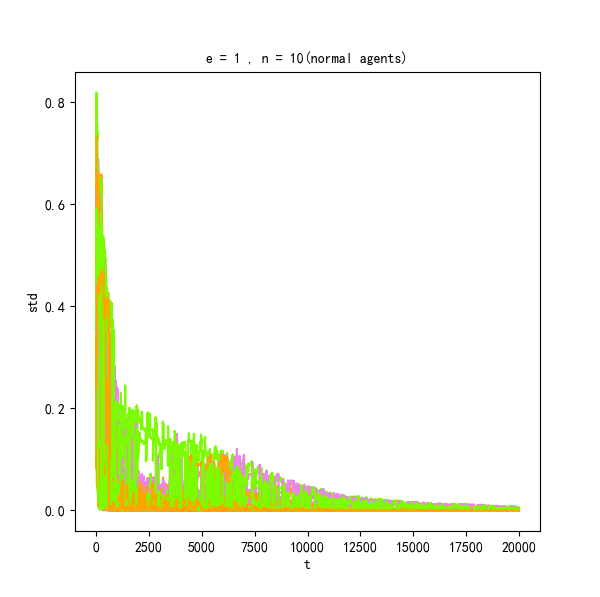}
		\subcaption{}
		\label{9-6}
	\end{minipage}
	\caption{Opinion evolution diagram with 10 active agents}
	\label{fig9}
\end{figure}

In Figure 9, according to subfigure 1, it can be found that there are three opinion clusters. As time goes by, the three opinion clusters are gradually merged into one, converging to the vicinity of the neutral opinion value. According to subfigure 2, it can be found that the opinion values of active nodes fluctuate around the neutral opinion, and the opinion values of some active nodes converge to the neutral opinion in a near gradient manner. By comparing subfigure 1 and Subfigure 2, it can be seen that active nodes can better guide the opinions of agents near the median to become more neutral, and free active nodes trend has a certain correspondence with the general nodes trend. The evolution of the network is shown in subfigure 3, and we find that almost all node opinions converge around 0 in the end. In addition, by comparing subfigure 4 and subfigure 5, it can be seen that before $t \approx 2000$, the degree of nodes in the network will fluctuate similarly with time, and after that, the degree of nodes in the network basically tends to be stable, that is to say, the topology of the network tends to be stable. As can be seen from subfigure 6, the standard deviation of neighbor of nodes fluctuates and converges over time, indicating that it is influenced by active nodes and the information sources of nodes are relatively diverse.

Further, from the micro level, we will specifically analyze the influence of the behavior of active agents on the formation of echo chamber effect. According to the subfigure 2 in figure 9, when $0 < t < 2000$, active nodes show transition fluctuation between heterogeneous opinions. For simplification, the transition states of active nodes are denoted as  ${h_1} =  - 0.25$ and  ${h_2} =  0.25$, and the corresponding influence range is  ${I_{{h_1}}} = [ - 0.5625, - 0.04],{I_{{h_2}}} = [0.04,0.5625]$.

First of all, we abstract the transition of an active agent's opinion as the reciprocal transformation between states and, that is, the opinion value of an active individual can only change back and forth between and, and no other opinion value will appear.

Now, consider the effect of the active node transition mechanism on a single agent’s near the neutral opinion. In an opinion interaction process, suppose there is an agent whose parameter is ${\mu _1},{\mu _2}$ and original opinion value is ${x_0}$, and the opinions ${h_1}$ and ${h_2}$ of active nodes are always within their trust threshold, then the following opinion update process will occur:
\begin{equation}
    \begin{array}{c}
{x_1} = (1 - {\mu _1} - {\mu _2}){x_0} + {\mu _1}{h_1} + {\mu _2}{h_2}\\
{x_2} = (1 - {\mu _1} - {\mu _2}){x_1} + {\mu _1}{h_1} + {\mu _2}{h_2}\\
 \vdots \\
{x_{n - 1}} = (1 - {\mu _1} - {\mu _2}){x_{n - 2}} + {\mu _1}{h_1} + {\mu _2}{h_2}\\
{x_n} = (1 - {\mu _1} - {\mu _2}){x_{n - 1}} + {\mu _1}{h_1} + {\mu _2}{h_2}
\end{array}
\end{equation}
We know by calculation that when $n\gg N$,$N \in \mathbb{Z}^+$:
\begin{equation}
    {x_n} \approx \frac{{{\mu _1}}}{{{\mu _1} + {\mu _2}}}{h_1} + \frac{{{\mu _2}}}{{{\mu _1} + {\mu _2}}}{h_2}
\end{equation}
The specific calculation process is as follows:

\begin{gather}
  {x_n} = (1 - {\mu _1} - {\mu _2}){x_{n - 1}} + {\mu _1}{h_1} + {\mu _2}{h_2} \\
   = (1 - {\mu _1} - {\mu _2})((1 - {\mu _1} - {\mu _2}){x_{n - 2}} + {\mu _1}{h_1} + {\mu _2}{h_2}) + {\mu _1}{h_1} + {\mu _2}{h_2} \\
   = {(1 - {\mu _1} - {\mu _2})^2}{x_{n - 2}} + {\mu _1}(1 + (1 - {\mu _1} - {\mu _2})){h_1} + {\mu _2}(1 + (1 - {\mu _1} - {\mu _2})){h_2} \\
                              \vdots  \\
   = {(1 - {\mu _1} - {\mu _2})^n}{x_0} + {\mu _1}(1 + (1 - {\mu _1} - {\mu _2}) +  \cdots  + {(1 - {\mu _1} - {\mu _2})^n}){h_1} +  \\
  {\text{  }}{\mu _2}(1 + (1 - {\mu _1} - {\mu _2}) +  \cdots  + {(1 - {\mu _1} - {\mu _2})^n}){h_2} \\
\end{gather}

When $n >  > N,N \in {\mathbb{Z}^ + }$:
$$\begin{gathered}
  {\mu _1}(1 + (1 - {\mu _1} - {\mu _2}) +  \cdots  + {(1 - {\mu _1} - {\mu _2})^n}) \hfill \\
   = {\mu _1} \times \frac{1}{{1 - (1 - {\mu _1} - {\mu _2})}} \hfill \\
   = \frac{{{\mu _1}}}{{{\mu _1} + {\mu _2}}} \hfill \\
\end{gathered} $$

Similarly,${\mu _2}(1 + (1 - {\mu _1} - {\mu _2}) +  \cdots  + {(1 - {\mu _1} - {\mu _2})^n}) = \frac{{{\mu _2}}}{{{\mu _1} + {\mu _2}}}$

Since $1 - {\mu _1} - {\mu _2} \in (0,1)$

When $n >  > N,N \in {\mathbb{Z}^ + }$: ${(1 - {\mu _1} - {\mu _2})^n} \to 0$

Then ${x_n} \approx \frac{{{\mu _1}}}{{{\mu _1} + {\mu _2}}}{h_1} + \frac{{{\mu _2}}}{{{\mu _1} + {\mu _2}}}{h_2}$

It can be seen that when N is sufficiently large, the opinions of nodes are the weighted average of the opinions of active nodes, and fluctuate in a certain range over time. From the above discussion, it can be seen that when nodes have high trust threshold, active agents can reduce the information narrowing caused by echo chamber effect by repeatedly providing heterogeneous information to nodes, so that opinions can converge to neutral opinions eventually. Therefore, on the whole, active agents have a better guiding and aggregating effect on the opinions near the neutral opinion.

Furthermore, we consider the effect of active nodes on nodes who are far from neutral. From Figure 9, it is not difficult to find that some opinions far from the neutral opinion converge to the neutral opinion with the convergence of opinions of active nodes. In fact, for active nodes with high opinion stubbornness, the fluctuation of opinion value is much smaller than active nodes with low opinion stubbornness. We then consider the effect of active nodes stubbornness on opinion far from neutral .

Next, the convergence parameters of active nodes are set as ${\mu _1} = {\mu _2} = 0.3$ and ${\mu _1} = {\mu _2} = 0.1$ respectively, and their stubbornness to their own opinions is 0.4 and 0.8 respectively. The specific experimental results are shown as follows:

\begin{figure}[htbp]
	\centering
	\begin{minipage}{0.32\linewidth}
		\centering
		\includegraphics[width=0.9\linewidth]{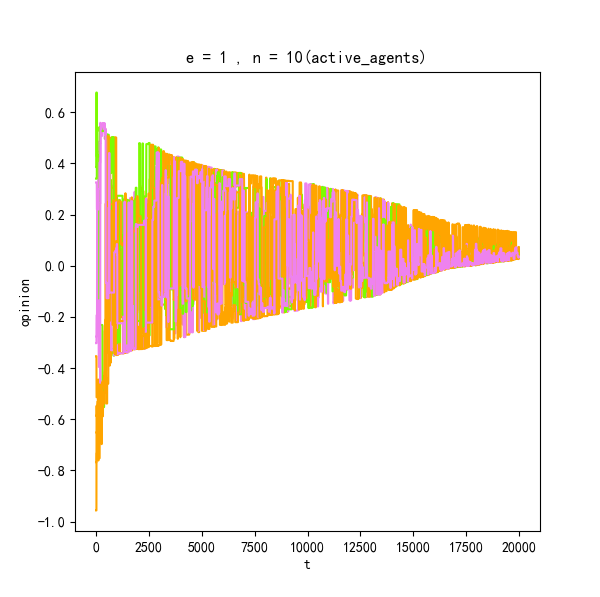}
		\subcaption{}
		\label{10-1}
	\end{minipage}
	\begin{minipage}{0.32\linewidth}
		\centering
		\includegraphics[width=0.9\linewidth]{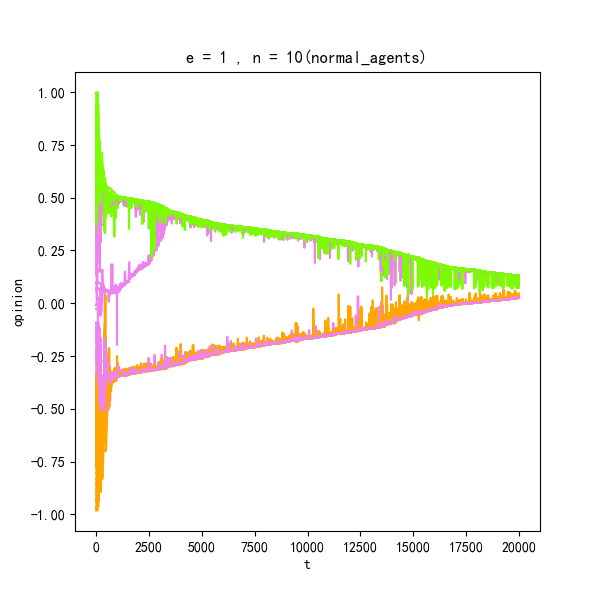}
		\subcaption{}
		\label{10-2}
	\end{minipage}
	\begin{minipage}{0.32\linewidth}
		\centering
		\includegraphics[width=0.9\linewidth]{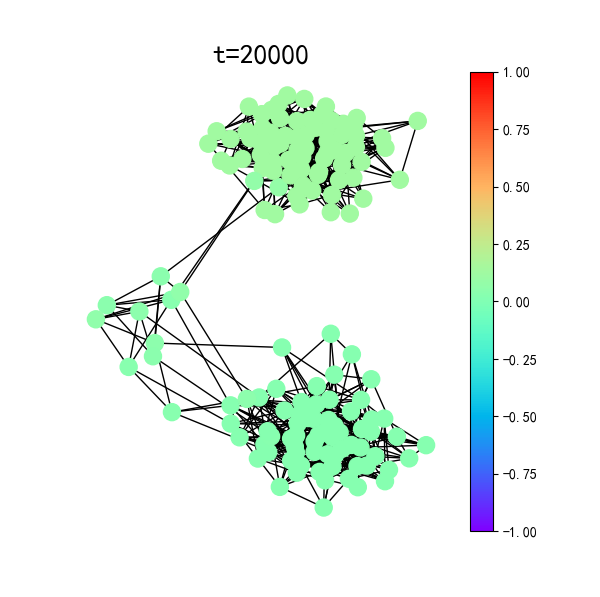}
		\subcaption{}
		\label{10-3}
	\end{minipage}
	\begin{minipage}{0.32\linewidth}
		\centering
		\includegraphics[width=0.9\linewidth]{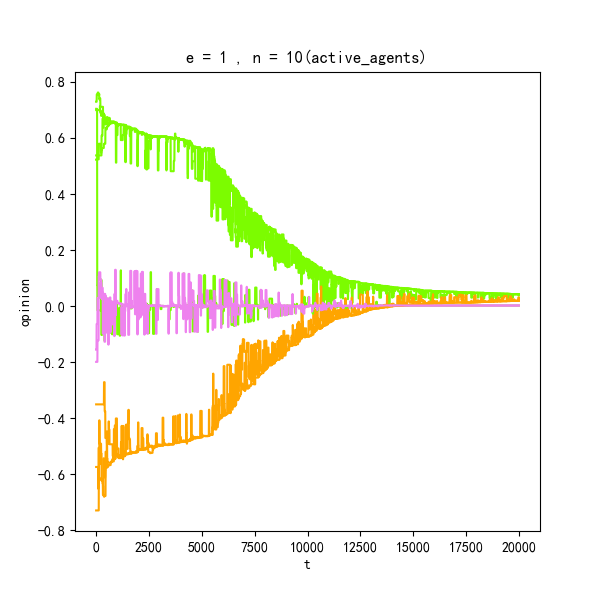}
		\subcaption{}
		\label{10-4}
	\end{minipage}
	\begin{minipage}{0.32\linewidth}
		\centering
		\includegraphics[width=0.9\linewidth]{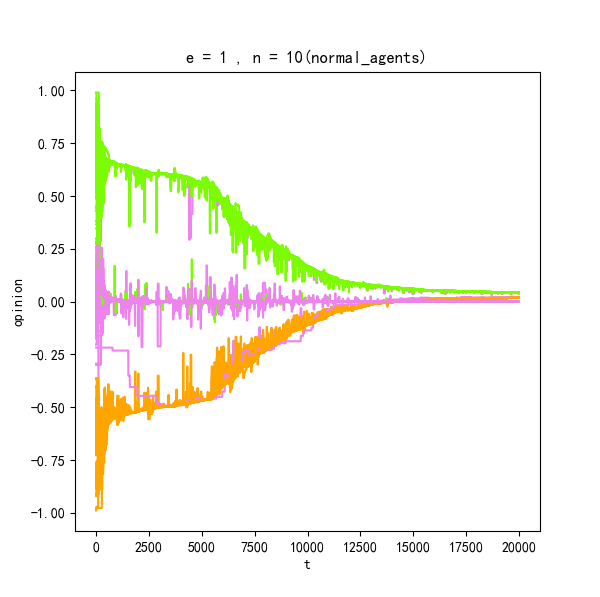}
		\subcaption{}
		\label{10-5}
	\end{minipage}
	\begin{minipage}{0.32\linewidth}
		\centering
		\includegraphics[width=0.9\linewidth]{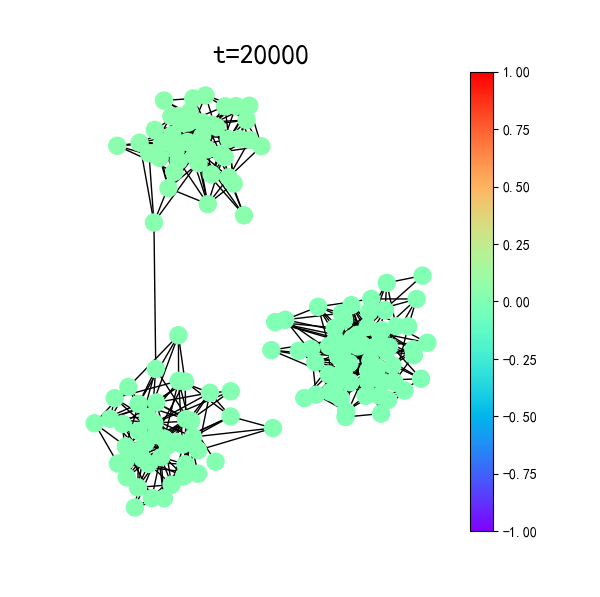}
		\subcaption{}
		\label{10-6}
	\end{minipage}
	\caption{Opinion evolution diagram with 10 active agents}
	\label{fig10}
\end{figure}
Figure 10 shows the model evolution results when the stubbornness of active nodes is 0.4 and 0.8 respectively. From the figure, it is not difficult to find that when active nodes with high stubbornness are not considered, active nodes with low stubbornness have limited ability to aggregate opinions and cannot well aggregate opinions near neutral opinions within a limited time. However, for the active nodes with high stubbornness, the opinions of the active nodes fluctuated less, and also aggregated the extremely dissociative nodes, making the final opinion near the neutral opinion.

In conclusion, active nodes' opinion updating behavior increases the contact probability with heterogeneous opinions, and its high trust threshold makes it easier for heterogeneous opinions to interact with each other, which makes it easier for active nodes' opinions to change to heterogeneous ones, and enables active nodes' opinions to fluctuate up and down in a transition pattern near neutral opinions. Finally, it can lead other opinions to converge to neutral. On the other hand, the existence of active nodes also makes some extreme opinions gradually converge to neutral opinions under the guidance of active opinions, thus breaking the echo chamber effect. From the perspective of agents, agents should not only contact with homogeneous opinions in the process of opinion interaction, but should take the initiative to contact with some heterogeneous opinions and keep an open and inclusive attitude towards the opinions of other agents. In this way, they can keep relatively rational in public opinion events, avoid excessive polarization of agent’s opinions and effectively avoid the generation of echo chamber effect.

\section{Conclusions}
In recent years, the corresponding existence of the echo chamber has brought some troubles to the governance of the network environment. The formation of the echo chamber has led to the emergence of more extreme and wrong opinions and damaged the network environment. Therefore, it is very important to study the formation mechanism of the echo chamber effect.

In this paper, the original D-W model is generalized to three dimensions, and corresponding improvements are made in trust threshold and interaction environment. From the perspective of social networks, it is found that although the convergence rate of opinions of social networks with different structures is different, heterogeneous group isolation will eventually occur, resulting in echo chamber effect. Although increasing the connectivity and stability of the network can reduce the isolation of heterogeneous groups, it cannot break the echo chamber effect. Therefore, we believe that under the natural evolution of opinions, no matter how the network structure is, echo chamber effect will eventually occur. On this basis, opinion leaders and active agents are introduced into the model. Opinion leaders have strong influence, but their scope of influence is limited. If the opinion value of opinion leaders remains unchanged, the purpose of opinion aggregation cannot be achieved. Only when opinion leaders gradually change their own opinions while guiding opinions within the scope of influence, and achieve the role of opinion guidance, can they finally break the echo chamber. For active agents, it increases the probability of contact with heterogeneous information. The diversity of information channels makes the phenomenon of information narrowing no longer happen. A certain number of active agents can eventually lead opinions to converge to neutral, thus breaking the echo chamber.

Compared with previous studies\cite{29}, the 3D D-W model in this paper is closer to the interaction of opinions in reality and has a faster convergence rate of opinions. In a practical sense, this paper considers the influence scope of opinion leaders' opinions, and believes that opinion leaders' opinions should not be unchanged, nor should they only express neutral opinions, so as to better show the guiding role of opinion leaders. On the other hand, active agents rarely appear in previous studies, and its practical significance is also obvious. Agents should be exposed to and tolerant of multiple opinions to enrich their own information sources and prevent the narrowing of opinion channels, which can effectively prevent the echo chamber effect.

Of course, there are some shortcomings in the study of this paper. First of all, the social network considered in this paper is the most basic network, and the network structure is also much simplified, there is a certain gap with the real social network. Secondly, although emotional factors are introduced into the model, the research on the psychological mechanism of agents is still not perfect. Finally, in the Internet era, agents can be active on multiple social platforms, so single-layer social networks have certain limitations. Therefore, in future study, we will focus on improving the shortcomings of the above three aspects.

\section*{Conflict of interest}

The authors declare that they have no conflict of interest.

\section*{Acknowledgment}
We would like to thank an anonymous reviewer for providing useful suggestions for improving the presentation of this work.

\bibliography{mybibfile}

\end{document}